\documentclass{article}

\usepackage{hyperref}

\usepackage[accepted]{icml2024}

\usepackage[utf8]{inputenc}
\usepackage{amsmath,amssymb,amsfonts,enumitem,color,algorithm,algorithmic,amsthm,bm}

\newcommand{\E}{\mathbb{E}}
\newcommand{\R}{\mathbb{R}}
\newcommand{\Aij}{A_{ij}}

\newcommand{\nmo}{n^{-\Omega(1)}}
\newcommand{\T}{\top}
\newcommand{\xcheck}{\check{x}}

\newcommand{\J}{\mathcal{J}}
\newcommand{\C}{\mathcal{C}}
\newcommand{\Lap}{\text{Lap}}
\newcommand{\argmax}{\text{arg max}}
\newcommand{\A}{\mathcal{A}}
\newcommand{\M}{\mathcal{M}}
\newcommand{\Stbl}{\operatorname{Stbl}}
\newcommand{\Oracle}{\textsc{Oracle}}
\newcommand{\ahat}{\widehat{a}}
\newcommand{\bhat}{\widehat{b}}
\newcommand{\what}{\widehat{w}}
\newcommand{\rhohat}{\widehat{\rho}}

\newtheorem{lemma}{Lemma}
\newtheorem{theorem}{Theorem}
\newtheorem{definition}{Definition}

\newtheorem{corollary}{Corollary}
\newtheorem{proposition}{Proposition}

\icmltitlerunning{Differentially private exact recovery for stochastic block models}

\begin{document}

\onecolumn

\icmltitle{Differentially private exact recovery for stochastic block models}



\icmlsetsymbol{equal}{*}

\begin{icmlauthorlist}
\icmlauthor{Dung Nguyen}{yyy}
\icmlauthor{Anil Vullikanti}{yyy}
\end{icmlauthorlist}

\icmlaffiliation{yyy}{Department of Computer Science and Biocomplexity Institute and Initiative, University of Virginia, Virginia, USA}

\icmlcorrespondingauthor{Dung Nguyen}{dungn@virginia.edu}

\icmlkeywords{Machine Learning, ICML}


\vskip 0.3in

\printAffiliations{} 

\setlength\parindent{0pt}

\begin{abstract}
Stochastic block models (SBMs) are a very commonly studied network model for community detection algorithms.
In the standard form of an SBM, the $n$ vertices (or nodes) of a graph are generally divided into multiple pre-determined communities (or clusters). 
Connections between pairs of vertices are generated randomly and independently with pre-defined probabilities, which depend on the communities containing the two nodes.
A fundamental problem in SBMs is the recovery of the community structure, and sharp information-theoretic bounds are known for recoverability for many versions of SBMs.

Our focus here is the recoverability problem in SBMs when the network is private.
Under the edge differential privacy model, we derive conditions for exact recoverability in three different versions of SBMs, namely Asymmetric SBM (when communities have non-uniform sizes), General Structure SBM (with outliers), and Censored SBM (with edge features).
Our private algorithms have polynomial running time w.r.t. the input graph's size, and match the recovery thresholds of the non-private setting when $\epsilon\rightarrow\infty$.
In contrast, the previous best results for recoverability in SBMs only hold for the symmetric case (equal size communities), and run in quasi-polynomial time, or in polynomial time with recovery thresholds being tight up to some constants from the non-private settings.
\end{abstract}

\section{Introduction}
\label{sec:intro}

A very common first step in the analysis of networked data in numerous applications, and unsupervised machine learning is  community detection or clustering, i.e., partitioning the network into ``well-connected'' communities, e.g.,~\cite{Blondel2008FastUO,girvan2002community,holland1983stochastic} (see survey by~\cite{fortunato2010community}).
There is limited theoretical understanding of community detection, since these notions are very problem-specific.
One exception is the stochastic block model (SBM)~\cite{holland1983stochastic}, a probabilistic generative model with well-defined communities, making them very amenable from a theoretical perspective.
Therefore, SBMs have been extensively studied in network science, and have become standard test benches for community detection algorithms. 

In the standard form of an SBM, the $n$ vertices (or nodes) of a graph are generally divided into multiple pre-determined communities (or clusters). 
Connections between pairs of vertices are generated randomly and independently with pre-defined probabilities, which depend on the communities containing the two nodes; the Erd\H{o}s-R\'enyi model is a special case of SBM with a single cluster.
The simplest type of SBM is a Binary Symmetric SBM (BSSBM)~\cite{abbe2015exact}, which consists of two communities with $n/2$ nodes each.
A pair of nodes within the same community are connected (forming an intra-cluster edge) with probability $p$, while a pair of nodes in the separate communities are connected with probability $q$ (forming an inter-cluster edge).
BSSBMs are quite restricted in terms of their structure, and many more complex SBMs have been developed to model more realistic networks, such as: multiple equal-sized clusters (Symmetric SBMs (SSBM)), 
allowing two clusters of unequal sizes (Binary Asymmetric SBM (BASBM))~\cite{hajek2016extensions},  combining them together and with the existence of outliers (General Structure SBM (GSSBM))~\cite{hajek2016achieving}, adding features such as weight labels (Censored SBM (CSBM))~\cite{hajek2016extensions}, and by introducing new assumptions such as degree distribution of vertices (Degree Corrected SBM), e.g.,~\cite{qin2013regularized}; see~\cite{lee2019review} for a survey on SBMs.

A fundamental problem is ``recovering'' the community structure from a given SBM, and this has spurred a very active area of research.
\emph{Exact recovery} is defined as when the probability that the community detection algorithm successfully recovers the ground-truth communities converges to $1$ when the size of the input graph (the number of vertices $n$) goes to infinity, with the probability space is over the randomness of the SBM process~\cite{abbe2015exact}. 
A celebrated result in this area is the exact recovery condition for BSSBM, namely that it is possible if  $p = a\log{n}/n$, $q = b\log{n}/n$ for constants $a$ and $b$, and $\sqrt{a} - \sqrt{b} \geq \sqrt{2}$. 
In some other regimes, for example, when $\sqrt{a} - \sqrt{b} < 2$ in the above setting, or when $p, q = \Theta(1/n)$, it has been shown that exact recovery is impossible.
When $\sqrt{a} - \sqrt{b} \geq \sqrt{2}$, exact recovery can be achieved by many different community detection algorithms, such as using a Maximum Likelihood Estimator (MLE), Spectral methods, or by Semi-definite programming (SDP)~\cite{boppana1987eigenvalues, mcsherry2001spectral,abbe2015exact,  massoulie2014community,gao2017achieving, hajek2016achieving, abbe2020entrywise, wang2020nearly}.  
Exact recovery has been studied for more complex SBMs, such as BASBM, GSSBM, and CSBM~\cite{hajek2016extensions, hajek2016achieving}, but the conditions are much more complex.
For instance, in BASBM, the threshold for exact recovery is when $\frac{a+b}{2} - \gamma + \frac{(1-2\rho)\tau}{2}\log\frac{\rho(\gamma + (1-2\rho)\tau)}{(1-\rho)(\gamma - (1-2\rho)\tau)} > 1$, where $\tau = (a-b)/(\log{a} - \log{b})$ and $\gamma = \sqrt{(1-2\rho)^2\tau^2 + 4\rho(1-\rho)ab}$~\cite{hajek2016extensions}.

Data privacy is a very significant concern in a number of applications.  
Differential Privacy (DP)~\cite{dwork2014algorithmic} has become a \emph{de facto} standard for privacy, due to its rigorous guarantees. 
DP algorithms guarantee that their outcomes will be similar probabilistically, measured by privacy parameters $\epsilon, \delta$, if the input is slightly modified. 
In context of networks and graph algorithms, two common privacy models have been considered, namely edge- and node-privacy~\cite{Kasiviswanathan:2013:AGN:2450206.2450232, blocki:itcs13,mulle2015privacy, nguyen2016detecting, qin2017generating, imola2021locally,blocki:itcs13}.
The edge-privacy model protects the existence and non-existence of an arbitrary edge in the input graph. In contrast, the node-privacy model provides protection to any node and its incident edges. 
Most work on private algorithms on community structures of graphs has focused on the edge-DP model, since the output contains nodes, e.g., densest subgraph~\cite{pmlr-v139-nguyen21i}, community detection~\cite{hehir2021consistency, pmlr-v162-mohamed22a, nguyen2016detecting}. 
Though the node-privacy model provides a stronger privacy guarantee, the edge-privacy model still provides meaningful protection in a number of applications.
A concrete example of the protection of edge-DP is against ``Link disclosure'' attack in social network analysis~\cite{zhou2008brief, kiranmayi2021review}, where users are modeled as nodes and social relationships between users are edges in the social graphs. For example, in the analysis of a communication graph that models the email interactions between students and faculty members in a university, in which the relationship ``who emails whom'' is considered sensitive, edge privacy can be applied to protect the sensitive links to be exposed~\cite{jiang2021applications}.
We refer readers to these studies~\cite{li2023private, jiang2021applications} for the motivation for edge-DP and its protection in many practical applications.

Community detection for SBMs under DP constraints (especially edge-DP) has been studied extensively in recent years~\cite{hehir2021consistency, pmlr-v162-mohamed22a, seif2023differentially, chen2023private, guo2023privacy}.
The first rigorous bound for recoverability in SBMs was established recently by~\cite{pmlr-v162-mohamed22a} for the special case of symmetric SBMs; for BSSBMs, they show that the condition for exact recovery is $\sqrt{a}-\sqrt{b} > \sqrt{2} \cdot \sqrt{1 + 3/2\epsilon}$, and that this can be extended to the case of $r$ communities.
\emph{The conditions for exact recovery in all other SBMs under differential privacy remain open.}

\subsection{Our contributions}

We study the exact recoverability problem in SBMs under the edge DP model--- this model is the natural model to consider (instead of node DP), since our goal is to output the community structure.
We consider three important extensions to the symmetric SBM model.

\begin{itemize}
      \item \emph{Binary Asymmetric} with unequal-sized clusters of size $\rho n$ and $(1-\rho)n$, for some constant $\rho\in[0, 0.5]$.
      \item \emph{Binary Censored}, in which edges of a graph $\mathcal{G}(n, p)$ (from the  Erd\H{o}s-R\'enyi model) are labeled as follows: an intra-cluster (or inter-cluster) edge has label $\bm{1}$ (or $\bm{-1}$, respectively) with probability $1-\xi$, and has the opposite label $\bm{-1}$ (or $\bm{1}$, respectively) with probability $\xi$, for some constant $\xi\in[0,0.5]$.
      \item \emph{General Structure}, consisting of multiple and possibly unequal clusters with outlier vertices that do not belong to any cluster.
      Each intra-cluster connection is generated with probability $p$, and all other connections are generated with probability $q$.
    \end{itemize}

\emph{We derive the first rigorous recoverability conditions in three extensions of SBMs under the edge DP model, and design polynomial time algorithms for recoverability}. 
Our results significantly extend the prior best theoretical result for recoverability in symmetric SBMs~\cite{pmlr-v162-mohamed22a}.

We establish the rigorous conditions of parameters for recoverability by sophisticated analysis of the stability of the Semi-definite program estimator on retrieving the true cluster mapping for each SBM variant.  
The condition involves both the graph model's parameters (edge probabilities $p$ and $q$ or edge label noisiness $\xi$), and the privacy parameters $\epsilon$ and $\delta$, to formalize the roles of different factors in the success of recovery.

The main difficulties are the design and analysis of a set of core conditions named $\C$-concentration for each model with three simultaneous properties: (1) being satisfied by a graph generated by the SBM with high probability (which refers to probability at least $1-1/n^c$ for a constant $c$, and abbreviated by w.h.p.), (2) persisting  under input graph perturbation up to $\log{n}$ edges, and (3) being sufficient to construct a dual certificate for the SDP Relaxation deterministically. The SBMs of interest are all significantly more complex than the symmetric SBMs, and differ substantially from each other. Each SBM requires a new design of $\C$-concentration, with little common analysis among them. 

We summarize the threshold in Table~\ref{table:summary-bounds}.
Finally, we design the first polynomial-time algorithms for community detection for SBMs with edge-DP for arbitrary small privacy parameter $\epsilon$, matching the exact recovery threshold in non-private settings when $\epsilon\rightarrow\infty$; \emph{none of the prior methods provide these guarantees.}

Our results advance the boundaries of recoverability with privacy on multiple variants of SBM.
Our focus here is on the theoretical foundations of the problem.

\begin{table*}[!ht]
    \centering
    \begin{tabular}{| c |c | c | c |}
      \hline
      Algorithm & SBM Model &  Recovery threshold & Running time \\
      \hline
      Non-private SDP & Binary Symmetric & $\sqrt{a} - \sqrt{b} \geq \sqrt{2}$ & $O(poly(n))$\\
      \hline
      MLE-Stability$^{(1)}$ & Binary Symmetric & $\sqrt{a} - \sqrt{b} \geq \sqrt{2}\sqrt{1 + 3/(2\epsilon)}$ & $O(exp(n))$\\
      SDP-Stability$^{(1)}$ & Binary Symmetric & $\sqrt{a} - \sqrt{b} \geq \sqrt{2}\sqrt{2 + 3/(2\epsilon)}$ & $n^{O(\log{n})}$\\
      RR + SDP$^{(1)}$ & Binary Symmetric & $\epsilon = \Omega(\log{n})$, $\sqrt{a} - \sqrt{b} > \sqrt{2} \times \frac{\sqrt{e^{\epsilon} + 1}}{\sqrt{e^{\epsilon} - 1}} + \frac{1}{\sqrt{e^{\epsilon} - 1}}$ & $O(poly(n))$ \\
      \cite{chen2023private} & Binary Symmetric & $\sqrt{a} - \sqrt{b} \geq 16, a -b \geq \frac{500^2}{\epsilon^2} + \frac{64}{\epsilon}^{\bm{(4)}}$ & $O(poly(n))$\\
      \cite{seif2023differentially} (MLE) & Binary Symmetric & $ \sqrt{a} - \sqrt{b} \geq \sqrt{2/\zeta}\sqrt{1 + 3/(2\epsilon)\Theta(\log{(a/b)})}$ & $O(exp(n))$ \\
      \cite{seif2023differentially} (SDP) & Binary Symmetric & $ \sqrt{a} - \sqrt{b} \geq \sqrt{2/\zeta}\sqrt{1 + 3/(\epsilon)\Theta(\log{(a/b)})}^{\bm{(5)}}$ & $n^{O(\log{n})}$ \\
      \hline
      $\M^{SDP}_{\Stbl\textsc{Fast}}$ \textbf{(Ours)} & Binary Symmetric$^{(2)}$ & $\sqrt{a} - \sqrt{b} \geq \sqrt{2}\sqrt{1 + 2/\epsilon}$ (Cor.~\ref{cor:threshold}) & $O(poly(n))$\\
 &&$\sqrt{a} - \sqrt{b(1+\log{\frac{a}{b}})} > \sqrt{c\log{\frac{a}{b}}/\epsilon}^{\bm{(6)}}$&\\
      \hline
       & Binary Asymmetric & Theorem~\ref{theorem:main-basbm} &$O(poly(n))$ \\
      $\M^{SDP}_{\Stbl\textsc{Fast}}$ \textbf{(Ours)} & Binary Censored & Theorem~\ref{threorem:main-cbsbm} & $O(poly(n))^{(3)}$\\
       & General Structure & Theorem~\ref{theorem:main-gssbm} & $O(poly(n))^{(3)}$\\

      \hline
    \end{tabular}
    \caption{Summary of the algorithms; SBM variants in the dense regime $p=a\log{n}/n, q=b\log{n}/n$; recovery thresholds on $a$, $b$ and $\epsilon$; time-complexity; at $\delta = n^{-2}$. For $\epsilon$-DP, both~\cite{pmlr-v162-mohamed22a,chen2023private} achieve exact recovery in exponential time. $(1)$ Algorithms of~\cite{pmlr-v162-mohamed22a}. $(2)$ The threshold derives from our BASBM with $\rho=1/2$ (Corollary~\ref{cor:threshold}). $(3)$ for Binary Censored and General Structure, $\M^{SDP}_{\Stbl\textsc{Fast}}$ takes $O(poly(n))$ if parameters $a$, $b$ and $\xi$ are known by the algorithms, otherwise we use $\M^{SDP}_{\Stbl}$ that takes $n^{O(\log{n})}$. $(4)$ is loosely equivalent to $\sqrt{a}-\sqrt{b} \geq 500/\epsilon + 8/\sqrt{\epsilon}$. $(5)\zeta$ denotes vertex sampling probability. $(6)$ when $\epsilon\rightarrow\infty$, this condition is roughly equivalent to $a > b$.}
  \label{table:summary-bounds}
\end{table*}


\subsection{Related work}

\cite{pmlr-v162-mohamed22a} are the first to derive rigorous bounds of exact recovery for SBMs under the edge-DP model. They proposed several approaches, notably applying the Stability mechanism on MLE and SDP estimator to achieve exact recovery in the symmetric settings with arbitrary small $\epsilon$ in $(\epsilon,\delta)$-DP, with exponential and quasi-polynomial time, respectively. Additionally, their proposed methods can achieve exact recovery in polynomial time but with the cost of $\epsilon = \Omega(\log{n})$, or in pure-DP with the cost of exponential time. Their best recovery threshold has the bound $\sqrt{a}-\sqrt{b}$ proportional to $1/\sqrt{\epsilon}$, and matches the non-private setting when $\epsilon\rightarrow\infty$.
\cite{seif2023differentially} improved the computational complexity of~\cite{pmlr-v162-mohamed22a} by using i.i.d.~vertex sampling with probability $\zeta$ and performing the Stability mechanisms on the sampled subgraph. After using private voting via the Laplace mechanism to classify unsampled nodes, they showed that exact recovery is achieved for both sampled and unsampled nodes, with the recovery thresholds increased by a factor of $1/\sqrt{\zeta}$. Their asymptotic running times remain unchanged, being exponential and quasi-polynomial for the MLE- and SDP-based mechanisms, respectively.
\cite{chen2023private} are the first to achieve exact recovery in polynomial time with any constant $\epsilon$. Their method relies on transforming the community detection problem into an optimization problem. They proved that once the optimization is strongly convex, the sensitivity of its solution is bounded, hence adding noises by the Gaussian mechanism provides privacy. This method achieves exact recovery in the BSSBM with the threshold being off from the non-private setting's threshold by some constant, and the bound $\sqrt{a} - \sqrt{b}$ being proportional to $1/\epsilon$ (see Table~\ref{table:summary-bounds}).~\cite{hehir2021consistency},~\cite{guo2023privacy},~\cite{ji2019differentially} studied the community detection in SBMs under DP, but did not focus on the exact recovery questions.

\section{Preliminaries}

\textbf{Stochastic Block Models.} The stochastic block model (SBM) is a family of random graph models in which a set of vertices $|V|=n$ is partitioned into $r$ clusters (communities): $C_1, \ldots, C_r$ plus some outliers that do not belong to any cluster. In binary forms ($r=2$), the clusters are represented by a vector $\sigma^*\in\{\pm 1\}^n$ where $\sigma^*_i = 1$ if vertex $i$ belongs to the first cluster and $\sigma^*_i = -1$ otherwise. When $r > 2$, clusters are represented by $r$ binary indicator vectors $\xi^*_1, \ldots, \xi^*_r \in \{0, 1\}^n$ where $\xi^*_k(i) = 1$ if vertex $i$ belongs to the cluster $k^{th}$ and $\xi^*_k(i) = 0$ otherwise. Connections between vertices are generated independently with probability $p$ if the endpoints are in the same cluster and with probability $q$ in other cases to form a set of edges $E$. In this paper, we focus on the dense regime, where $p = a\log{n}/n$ and $q = b\log{n}/n$ for some constants $a \geq b > 0$, of the following three variants of SBMs:

\begin{enumerate}
      \item \emph{Binary Asymmetric (BASBM)}: An SBM with $r = 2$ in which the first cluster contains $\lfloor n\rho \rfloor$ and the second one contains $\lceil n(1-\rho) \rceil$ vertices for some constant $\rho\in[0, 0.5]$. 
      \item \emph{Binary Censored (BCSBM)}: An SBM with $r=2$ and $p = q$. In other words, edges are generated by an Erdos-Renyi model $\mathcal{G}(n, p)$. Each edge $(i, j)$ has label $L_{ij}\in\{\pm 1\}$ independently drawn from the distribution: $P_{L_{ij}} = (1-\xi)\bm{1}_{L_{ij} =\sigma^*_i\sigma^*_j} + \xi\bm{1}_{L_{ij}=-\sigma^*_i\sigma^*_j}$.
      \item \emph{General Structure (GSSBM)}: An SBM with $r > 2$, where the $k^{th}$ cluster $C_k$ has size $K_k = \rho_kn$ and $\rho_1 \geq \ldots \geq \rho_r > 0$; and $n-\sum_{k\in[r]}K_k$ outliers. We use $k=0$ (e.g., in $C_0, K_0$) to refer to the outliers, but the edges among them are not considered intra-cluster.
  
\end{enumerate}

\textbf{Exact recovery.} Given the input graph $G = (V, E)$ as described above and an algorithm $\A$, \emph{exact recovery} means that $\A$ outputs the ground-truth cluster vector $\sigma^*$ up to a permutation w.h.p., i.e., with probability tending to $1$ when $n$ goes to infinity: $\Pr[\A(G) \neq \pm \sigma^*] = o(1)$. For $r>2$, we often use the cluster matrix: $\Pr[\A(G) \neq Z^*] = o(1)$, where $Z^*=\sum_{k\in[r]}\xi^*_k(\xi^*_k)^T$.

Exact recovery is not always possible, for example, in the \emph{sparse regime} $p,q = \Omega(1/n)$. In the symmetric SBM, exact recovery is possible in the \emph{dense regime} if and only $\sqrt{a} - \sqrt{b} \geq \sqrt{r}$. For Binary Asymmetric SBM, the threshold for exact recovery is when $\frac{a+b}{2} - \gamma + \frac{(1-2\rho)\tau}{2}\log\frac{\rho(\gamma + (1-2\rho)\tau)}{(1-\rho)(\gamma - (1-2\rho)\tau)} > 1$, where $\tau = (a-b)/(\log{a} - \log{b})$ and $\gamma = \sqrt{(1-2\rho)^2\tau^2 + 4\rho(1-\rho)ab}$. In the case of the Binary Censored model, the threshold is $a(\sqrt{1-\xi}-\sqrt{\xi})^2 > 1$. The condition for exact recovery in the General Structure is stated in~\cite{hajek2016extensions}.

\textbf{Privacy model.} We consider the exact recovery problem under the edge-privacy model~\cite{karwa2011private}. A randomized algorithm $\A$ is $(\epsilon,\delta)$-differentially private (DP) if for any pair of neighbor graphs that differ by exact one edge (denoted as $G\sim G'$), $\Pr[\A(G)\in S] \leq e^\epsilon\Pr[A(G')\in S] + \delta$ for any $S\subseteq Range(A)$. 

\textbf{Stability mechanism.} Given a function $f$, an input graph $G$ is called $c$-stable if $\forall G': dist(G, G') \leq c$, we have $f(G) = f(G')$, where $dist$ denotes the Hamming distance. A graph is unstable if it is $0$-stable. The \emph{distance to instability} of input $G$ on $f$, denoted by $d_f(G)$ is the shortest distance to reach to an unstable graph $G'$ from $G$. The $f$-based Stability mechanism, as defined in Algorithm~\ref{alg:stability-mechanism}, is $(\epsilon,\delta)$-DP~\cite{dwork2014algorithmic, pmlr-v162-mohamed22a} (The formal statement and proof are presented in Theorem~\ref{theorem:privacy}). It first calculate $G$'s private distance to instability, by adding a Laplacian noise with magnitude $1/\epsilon$ ($\Lap(b)$ denotes the Laplace distribution with PDF $\Lap(x|b) = 1/(2b)\exp(-|x|/b)$), as the distance to instability always have sensitivity of $1$. Line $5$ of the algorithm returns an undefined output ($\perp$) which keeps the privacy analysis simple, but in practice can be replaced by returning a random output. 
\begin{algorithm}
  \caption{$\M^{f}_{\Stbl}(G)$: Stability Mechanism}
  \label{alg:stability-mechanism}
  \begin{algorithmic}[1]
    \STATE $\tilde{d}_f(G) \leftarrow d_f(G) + \Lap(1/\epsilon)$
    \IF{$\tilde{d}_f(G)  > \frac{\log{1/\delta}}{\epsilon}$}
    \STATE Output $f(G)$
    \ELSE
    \STATE Output $\perp$ 
    \ENDIF
  \end{algorithmic}
\end{algorithm}

\textbf{Utility of Stability mechanism.} The utility of the Stability mechanism dictates its ability to achieve exact recovery. Let $f$ be any community detection algorithm that outputs the communities given an input $G$. Lemma~\ref{lemma:prelim-stability-mechanism} states that with appropriate selections of $\delta$, if an input graph $G$ is $O(\log{n})$-stable under $f$, then w.h.p. (with probability at least $1-n^{-\Omega(1)}$), the mechanism returns $f(G)$. We extensively utilize this property in our analyses. Hereafter, for simplicity, we will refer to $O(\log{n})$-stable (or distance up to $O(\log{n})$) as $\log{n}$-stable (or distance up to $\log{n}$).

\begin{lemma}
  \label{lemma:prelim-stability-mechanism}
  (Full proof in Lemma~\ref{lemma:stability-mechanism})
 The $f$-based Stability mechanism with $\delta = n^{-c}$ has $\Pr[M^f(G) \neq f(G)] \leq n^{-k_1} + n^{-k_2}$, if a graph $G$ is $ \frac{c + k_1}{\epsilon}\log{n} $-stable under function $f$ with probability at least $1-n^{-k_2}$. When $k_1, k_2 = \Omega(1)$, $\Pr[\M^f_{\Stbl} \neq f(G)] \leq n^{-\Omega(1)}$.
\end{lemma}

\textbf{Problem statement.} Given an input graph $G=(E, V)$, with $|V| = n$, assumed to generated by an SBM with a fixed but unknown cluster vector $\sigma^*$ (or community matrix $Y^*$ in case of GSSBM), design an algorithm $\A$ such that: (1) $\Pr[\A(G)\neq \pm \sigma^*] = o(1)$ (exact recovery property) and (2) $\A$ is $(\epsilon,\delta)$-DP.


\section{Exact Recovery under Differential Privacy}
\label{sec:sdp-based-stability-mech}
In non-private settings, using Semidefinite programming (SDP) to estimate the clusters is an established method for community detection and exact recovery. We apply the SDP estimator with the Stability mechanism to design our algorithms for differentially private exact recovery.

In this Section, we will present an inefficient algorithm by a direct application of the Stability mechanism to output a private SDP estimator with privacy. We show in this section a high-level overview of our analyses by introducing the generic conditions and analysis routines to guarantee exact recovery for all three targeted SBMs above. In Section~\ref{sec:stability-analyses}, we will present the custom-built analyses for each of the three models to fulfill the proof. 
Finally, in Section~\ref{sec:poly-alg}, we will show how to design a novel polynomial-time algorithm that still retains all privacy and utility guarantees. Due to space limits, we leave the complete proofs in the Appendix.

There are two main steps in designing the algorithms.
The first step is to define the SDP relaxation of the SBMs based on the input graph's adjacency matrix $A$. The second step is to treat the optimal solution of the SDP as a function (denoted as $SDP(G)$) and apply it in the role of $f$ in Algorithm~\ref{alg:stability-mechanism}.
Here is the SDP for the BASBM model:

\begin{minipage}[t]{0.45\linewidth}%
  \vspace{-0.215in}
  \begin{align}
    \max_\sigma&\sum_{ij}A_{ij}\sigma_i\sigma_j \\
    \text{s.t. } &\sigma_i\in\{\pm 1\}, i\in[n] \nonumber \\
               &o^T\bm{1} = n(2\rho-1) \nonumber
  \end{align}
\end{minipage}%
\hfill\vline\hfill
\begin{minipage}[t]{0.53\linewidth}%
  \vspace{-0.215in}
  \begin{align}
    \widehat{Y}_{SDP} =& \argmax_Y \langle A , Y \rangle \\
    \text{s.t. } &Y \succcurlyeq 0\nonumber\\
                       &Y_{ii} =1, i\in[n]\nonumber\\
                       \langle J, Y \rangle &= (n(2\rho-1))^2\nonumber
  \end{align}
\end{minipage}%

Let $\sigma$ be our estimator vector with $\sigma_i = 1$ if $i$ is in the first cluster and $\sigma_i = -1$ otherwise. The optimization problem (1) maximizes the differences between intra-cluster edges and inter-cluster edges while fixing the size of the two clusters (the last constraint), namely Maximum Likelihood Estimator (MLE). Solving this is NP-hard. We relax the problem into the SDP optimization (2) as follows: let $Y = \sigma\sigma^T$, $\sigma=\pm 1$ will be transformed to $Y_{ii} = 1$, and the size constraint will be equivalent to $\langle J, Y \rangle = (n(2\rho-1))^2$. All feasible matrices are rank-one positive semi-definite, which is then relaxed to $Y\succcurlyeq 0$. Depending on the unique specifications of each SBM, their equivalent SDPs are vastly different. Further details are presented in~\cite{hajek2016extensions}.

\textbf{Applying the Stability mechanism.} After defining the SDP (and the function $SDP(G)$, which denotes the optimal solution), we can directly apply it to the Stability mechanism by substituting $f(G)$ by $SDP(G)$ to form an $SDP$-based Stability mechanism. We note that for any function $f$, the $\M^f_{\Stbl}$ is $(\epsilon,\delta)$-DP. Theorem~\ref{theorem:privacy} formalizes the privacy analysis of the mechanism. \emph{Exact recovery} requires extensive analysis of SDP under the effect of the mechanism. To achieve exact recovery, $SDP(G)$ must be $\log{n}$-stable w.h.p., due to Lemma~\ref{lemma:prelim-stability-mechanism}. Though~\cite{pmlr-v162-mohamed22a} shows the analysis for the symmetric SBMs, and it cannot be extended easily to cover any of the three models above, due to their different SDP formulas, as well as different criteria of the certificates of the optimality of the SDPs. We next present the generic routine, which consists of the three main steps below, to prove the $\log{n}$-stability property for a generic SBM, and then provide a specific procedure for each of them in Section~\ref{sec:stability-analyses}. 

\textbf{Generic analysis of $\bm{\log{n}}$-stability of $\bm{SDP(G)}$} is centered around the concept of $\C$-concentration, parameterized by a tuple of constants $\C$. It is a set of special conditions constructed from the specifications of the assumed SBM. For example, $\Vert A-\E[A] \Vert\leq c\sqrt{\log{n}}$ is a condition parameterized by a constant $c$, where expected value $\E[A]$ of the adjacency matrix $A$ is dictated by the SBM. A graph $G$ is called $\C$-concentrated when it satisfies all conditions under $\C$. Each SBM, and its equivalent SDP, has a distinct $\C$-concentration. To prove that $SDP(G)$ is $\log{n}$-stable, we have to complete the following steps:

\textbf{Step 1. $\C$-concentration w.h.p.} Over the randomness of the generation process of an SBM, we prove that a random graph $G$ is $\C$-concentrated w.h.p.. It often requires each condition being satisfied w.h.p. and taking union bound on all conditions. For example, we prove that the condition $\Vert A-\E[A] \Vert\leq c\sqrt{\log{n}}$ be satisfied w.h.p. for a graph generated by the SBM. This step also sets the restrictions on the constants $\C$, e.g., $c$ in the above condition must be a positive constant, which will determine the thresholds for the exact recovery. It is formalized by the following proposition:

\begin{proposition}
  \label{prop:concen-high-prob}
	Given graph $G$ generated by an SBM of specific settings and its respective $\C$-concentration conditions, there exists some tuples of constants $\C=\{c_1,c_2,\ldots\}$ such that $G$ is $\C$-concentrated with probability at least $1 - \nmo$.
\end{proposition}

\textbf{Step 2. $\C$-concentration persists under up to $\bm{\log{n}}$ edges modifying.} By proving the following proposition, we show that modifying $\log{n}$ edges of a $\C$-concentrated graph $G$ does not destroy its concentration properties (up to a new tuple $\C'$ as long as $\C'$ satisfies the restriction in Step 1). For example, with the condition $\Vert A-\E[A] \Vert\leq c\sqrt{\log{n}}$, modifying up to $\log{n}$ edges of the input graph $G$ flips at most $2\log{n}$ bits of $A$ (due to the symmetry of $A$). For the condition to persist, we need a new constant $c' = c + 2$. Since $c' > 0$, it is a valid constant of $\C$.

\begin{proposition}
  \label{prop:concen-persist}
	If $G$ is $\C$-concentrated then for every graph $G': d(G, G') < c\log{n}$, i.e., $G'$ can be constructed by flipping at most $c\log{n}$ connections of $G$, $G'$ is $\C'$-concentrated, where $\C'$ is a valid tuple of constants depending only on $\C$, $c$, and the SBM's constant parameters.
\end{proposition}

\textbf{Step 3. $\C$-concentration implies SDP optimality at the ground-truth.} In this step, we will construct a (deterministic) dual certificate for the SDP Relaxation using the $\C$-concentration's conditions. Once we do that, we can confirm that the ground-truth cluster matrix ($Y^*$ or $Z^*$) is the unique optimal solution of the SDP Relaxation. In other words, when $\C$-concentration holds, $SDP(G)$ always outputs the ground-truth clusters. Lemma~\ref{lemma:binary-cert-main} states the certificates for the BASBM, i.e., the ground-truth community matrix $Y^*$ is the unique and optimal solution of the SDP given appropriate $D^*$ and $S^*$. \cite{hajek2016extensions} shows that $D^*$ and $S^*$ exist w.h.p. for the non-private setting. In our design, we show that from $\C$-concentration's conditions (such as $\Vert A-\E[A] \Vert\leq c\sqrt{\log{n}}$), we can always, i.e., with probability $1$, construct $D^*$ and $S^*$ that satisfies Lemma~\ref{lemma:binary-cert-main}. Applying the lemma, we show that $\C$-concentration implies that the SDP has the optimal solution at the exact ground-truth.

\begin{lemma} (Lemma 3 of~\cite{hajek2016extensions})
	\label{lemma:binary-cert-main}
	Suppose there exist $D^* = diag\{d^*_i\}$ and $\lambda^*\in\mathbb{R}$ such that $S^* = D^* -A +\lambda^*J$ satisfies $S^*\curlyeqsucc 0, \lambda_2(S^*) > 0$ and $S^*\sigma^* = 0$.
	Then $Y^*$ is the unique solution of the program $SDP(G)$.
\end{lemma}

\begin{proposition}
  \label{prop:concen-optimal}
	If $G$ is $\C$-concentrated under an SBM, then $SDP(G) = Y^* (\text{or }Z^*)$, i.e, the optimal solution $SDP(G)$ is the ground-truth cluster matrix $Y^*$ (or $Z^*$).
\end{proposition}

\textbf{Exact recovery.} Step 2 \& 3 guarantee that when an input graph $G$ is $\C$-concentrated under an SBM with specific settings, the function $SDP(G)$ is $\log{n}$-stable. Since $G$ is $\C$-concentrated, $SDP(G)$ outputs the ground-truth $Y^*$, or $SDP(G) = Y^*$, by Proposition~\ref{prop:concen-optimal}. Any graph $G'$ created by flipping up to $\log{n}$ edges of $G$ is $\C'$-concentrated, by Proposition~\ref{prop:concen-persist}. Now, applying Proposition~\ref{prop:concen-optimal} for $G'$, we have $SDP(G) = SDP(G') = Y^*$. In other words, $G$ is $\log{n}$-stable under the function $SDP$.

Finally, Step 1 shows that $\C$-concentration happens w.h.p. for an arbitrary graph generated by the SBM. In other words, if $G$ is generated by an SBM with appropriate parameters, $G$ is $\log{n}$-stable under the function $SDP$ w.h.p.. Applying Lemma~\ref{lemma:prelim-stability-mechanism}, substituting $f$ by $SDP$, $\Pr[\M^{SDP}_{\Stbl}(G) \neq Y^*(\text{or } Z^*)] \leq \nmo$, or the ground-truth clusters are recovered w.h.p..

\section{Stability analyses for SBMs}
\label{sec:stability-analyses}
In this Section, we develop the specific analysis for each of the three SBMs based on the generic routine in Section~\ref{sec:sdp-based-stability-mech}.

\subsection{Binary Asymmetric}

The SDP Relaxation of the Binary Asymmetric model is shown in Section~\ref{sec:sdp-based-stability-mech}. Recall that in BASBM, two clusters have size $\rho n$ and $(1-\rho)n$, $p = a\log{n}/n$, $b = b\log{n}/n$. We define the $\C$-concentration as follows:

\begin{definition}
	$G$ is called $\C = (c_1, c_2, c_3, c_4)$-concentrated, in which $c_i > 0$, if $G$ satisfies all $4$ conditions:
	\begin{itemize}
		\item $\Vert A - \E[A]\Vert_2 \leq c_1\sqrt{\log{n}}$
		\item $\xcheck^\T D^*\xcheck + \J(\xcheck) > c_2\log{n} $
		\item $\Vert (D^* - \E[D^*])\xcheck\Vert_2 \leq c_3\sqrt{\log{n}}$
		\item $d^*_i \geq c_4\log{n}$ for every $i\in[n]$;
	\end{itemize}
in which 
$\tau = \frac{a-b}{\log{a}-\log{b}}$,
$\lambda^* = \tau\log{n}/n$,
$d^*\in\mathbb{R}^n: d^*_i = \sum_{j=1}^n\Aij\sigma^*_i\sigma^*_j - \lambda^*(2K - n)\sigma^*_i$,
$D^* = diag\{d^*\}$,
$I$ is the identity matrix,
$J$ is the all-one matrix,
$\J(x) = \left(\lambda^* - \frac{p+q}{2}\right)x^\T Jx$,
$\xcheck = \arg \max_{x}x^\T Jx$ subject to $\Vert x \Vert_2 = 1$ and $\langle x, \sigma^* \rangle = 0$,
$\tilde{h}(x) = a\rho + b(1-\rho) - \sqrt{(\tau(1-2\rho) - x)^2 + 4\rho(1-\rho)ab} + \frac{\tau(1-2\rho) - x}{2}\log{\frac{\rho b}{(1-\rho)a}} $
.
\end{definition}

\textbf{Assumptions of the parameters.} These assumptions will be used in our analyses to derive the conditions in which exact recovery under DP is feasible.

The exists a constant $c > 0$ that for $\tilde{h}(x)$ defined above:
\begin{align}
	\tilde{h}(c) = 1 + \Omega(1)
	\label{eq:main-assumption-h}
\end{align}

\begin{lemma}
	\label{lemma:main-concen-high-prob}
    (Complete proof in Lemma~\ref{lemma:concen-high-prob})
	A graph $G$ generated by a Binary Asymmetric SBM for some constant $\rho\in[0,0.5]$ and with $p = \frac{a\log{n}}{n}$ and $q = \frac{b\log{n}}{n}$, there exists some tuples of constants $\C$ that $G$ is $\C$-concentrated with probability at least $1 - \nmo$.
\end{lemma}

\textbf{Proof sketch.} The adjacency matrix $A$ is a random matrix where each cell is drawn i.i.d with probability of being $1$ is $\log{n}/n$. The first condition follows directly from Lemma~\ref{lemma:adj-mat-concen}, saying with these properties, $A$ is usually not far from its expectation $\E[A]$. The second and third conditions are highly complex, and we have to decompose them into sub-components for further analysis. Notice that $D^*$, by its definition, is a random matrix computed from $A$. Therefore we can treat $\xcheck^TD^*\xcheck$ and $\Vert (D^* - \E[D^*]) \xcheck\Vert$ as functions that take independent variables as input, and apply Talagrand inequality (Lemma~\ref{lemma:talagrand}) that says the output of $1$-Lipschitz convex functions is not far from its expected values. They are then transformed to expressions of $\log{n}$ or $\sqrt{\log{n}}$ as we see in the r.h.s. of the conditions. With that, the remaining things to do is to prove the functions of interest are (1) convex and (2) Liptschitz continuous. The fourth condition involves $d^*_i$, or more specifically, $\sum_jA_{ij}\sigma^*_i\sigma^*_j$. This quantity is equal to the difference between two Binomial distributions with different numbers of trials and probabilities of success, and can be analyzed by Lemma~\ref{lemma:binom-diff}. The final detail of the proof is the tail bound of Lemma~\ref{lemma:binom-diff} refers to the assumption in Equation~\ref{eq:main-assumption-h}, so that the tail probability is set to less than $n^{-1-\Omega(1)}$ for each $d^*_i$.

\begin{lemma}
	\label{lemma:main-concen-stable}
    (Complete proof in Lemma~\ref{lemma:concen-stable})
	If $G$ is $\C=(c_1, c_2, c_3, c_4)$-concentrated then for every graph $G': d(G, G') < \frac{c}{\epsilon}\log{n}$, i.e., $G'$ can be constructed by flipping at most $\frac{c}{\epsilon}\log{n}$ connections of $G$, $G'$ is $\C' = (c_1', c_2', c_3', c_4')$-concentrated, where $c_1' = c_1 + \sqrt{2c/\epsilon}, c_2' = c_2-c/\epsilon, c_3' = c_3+\sqrt{2c(1-\rho)/\epsilon\rho}, c_4' = c_4-c/\epsilon$.
\end{lemma}

\textbf{Proof sketch.}
The general idea is to construct worst-case scenarios of quantities in the l.h.s. of the conditions when adding/removing $\log{n}$ edges and edit $\C$ accordingly such that all four conditions still hold for the new graph. For example, we construct the adjacency matrix $A'$ obtained by modifying $\log{n}$ entries of $A$. Assume $\E[A] = \E[A']$ because they are assumed to be generated by the same SBM, we can apply the triangle inequality property of $\ell$-norm to prove that the l.h.s. will be increased by at most $\Theta{\sqrt{\log{n}}}$. We then adjust $c'$ by the same constant to complete the proof for the first condition. For the second condition, notice that $D^*$ will change by edge modifying. Let $\Delta={D'}^*-D^*$, with ${D'}^*$ being the same quantity as $D^*$ but in $G'$. We can prove that $\Delta$ is also diagonal, with the sum of its entries being at most $\Theta(\log{n})$, and its largest eigenvalue is $\Theta(\log{n})$. Then the difference after modifying $\log{n}$ edges is reduced to the quantity $\xcheck^T\Delta\xcheck \leq \lambda_{max}(\Delta)\Vert\xcheck\Vert^2_2\leq\Theta(\log{n})$, which means we only need to alter $c_2$ by some small constant for the condition to persist. For the third condition, we reduce the change in the l.h.s. of the condition by triangle inequality to $\Vert({D'}^* - D^*)\xcheck\Vert$. Expand this quantity by expanding $D^*$ and ${D'}^*$ by their definitions to expressions of $A_{ij}$ and $A'_{ij}$, we can reduce it to the $F$-norm of $A-A'$, which is no greater than $\Theta({\log{n}}))$. Since the r.h.s. of the third condition is also $\Theta(\sqrt{\log{n}})$, we only need to change $c_3$ by some small constant for it to hold. For the fourth condition, for each $i$, the difference quantity is bounded by $\sum_j (A'_{ij}-A_{ij})\sigma^*_i\sigma^*_j$. Because $A$ and $A'$ differ by at most $\Theta(\log{n})$ entries, it is clear that changing the constant $c_4$ a bit will cover the changes.

\begin{lemma}
	\label{lemma:main-sdp-optimal}
    (Complete proof in Lemma~\ref{lemma:sdp-optimal})
	If $G$ is $\C$-concentrated, then $SDP(G) = Y^*$, i.e., the SDP's optimal solution is the ground-truth cluster matrix $Y^* = \sigma^*\sigma^{*T}$.
\end{lemma}

\textbf{Proof sketch.}
We will construct a (deterministic) dual certificate by the conditions of $\C$-concentration. Lemma~\ref{lemma:binary-cert} states that $SDP(G) = Y^*$, if we can construct a matrix $S^* \overset{def}{=}D^* - A + \lambda^*J$ satisfies (1) $S^* \succcurlyeq 0$, (2) $\lambda_2(S^*)>0$, and (3) $S^*\sigma^* = 0$. The condition (3) is easy to verify by expanding the definition of $S^*$. Because of this, proving $\inf_{x\perp \sigma^*, \Vert x \Vert = 1} x^TS^*x > 0$ is sufficient to satisfy all remaining conditions (1) and (2), since all feasible $x$ plus $\sigma^*$ will include a basis for the whole space, which means $\forall y: y^TS^*y \geq 0$ ($S^* \curlyeqsucc 0$) and the solution set of $S^*y = 0$ has only $1$ dimension ($\lambda_2(S^*) > 0$).

To prove $x^TS^*x > 0$, we have to utilize all four conditions of $\C$-concentration.
First we expand $x^TS^*x = x^TD^*x - x^TAx + x^T\lambda^*Jx$. Adding $x^T\E[A]x - x^T\E[A]x$ to the r.h.s. and grouping $x^TAx - x^T\E[A]x$, we can use the \textbf{first condition} to bound this quantity by $c_1\sqrt{\log{n}}$. We next expand the remaining $\E[A] = \frac{p-q}{2}Y^* + \frac{p+q}{2}J - pI$, and reduce the equation to $x^TS^*x \geq  p - c_1\sqrt{\log{n}} + x^TD^*x + (\lambda^* - \frac{p+q}{2})x^TJx$.

Let $t(x) = x^TD^*x + (\lambda^* - \frac{p+q}{2})x^TJx$. 
We define $E = span(\xcheck, \sigma^*)$. Any $y: y\perp \sigma^*, \Vert y \Vert_2 = 1$ can be represented as $y = \beta\xcheck + \sqrt{1-\beta^2}x$ for $x\in\{x: x \perp E, \Vert x \Vert_2 = 1\}$ and $\beta\in[0, 1]$. For all $x: x\perp \sigma^*, \Vert x \Vert_2 = 1$:

$\inf_{x} t(x) = \inf_{x,\beta\in[0,1]}t(\beta\xcheck + \sqrt{1-\beta^2}x)$:
	\begin{align}
      \geq &\inf_{\beta\in{0,1}}(\beta^2(\xcheck^TD^*\xcheck + \J(\xcheck)) \nonumber\\
      + &\inf_{x, \beta\in[0,1]}(2\beta\sqrt{1-\beta^2}x^TD^*\xcheck + (1-\beta^2)x^TD^*x) \nonumber\\
\geq &\inf_{\beta\in{0,1}}(\beta^2(\xcheck^TD^*\xcheck + \J(\xcheck)) \nonumber\\ &+ (1-\beta^2)c_4\log{n}) - c_3\sqrt{\log{n}}\nonumber\\
		                                                 \geq&\frac{1}{2}\min\{c_2, c_4\}\log{n} - c_3\sqrt{\log{n}},.
	\end{align}
where in the second last inequality, we apply the \textbf{second condition, third condition, fourth condition} to bound $\xcheck^\T D^*\xcheck + \J(\xcheck)$, $\Vert (D^* - \E[D^*])\xcheck\Vert_2$, $x^TD^*x$, respectively. Applying this result to  $x^TS^*x \geq  p - c_1\sqrt{\log{n}} + x^TD^*x + (\lambda^* - \frac{p+q}{2})x^TJx \geq\frac{1}{2}\min\{c_2, c_4\}\log{n} - (c_1+c_3)\sqrt{\log{n}} + p > 0$, when $n$ is large enough and the Theorem follows.

Finally, by the arguments in Section~\ref{sec:sdp-based-stability-mech}, we show that $\M_{\Stbl}^{SDP}$ achieves exact recovery as follows:

\begin{theorem}
   \label{theorem:main-basbm}
   (Complete proof in Theorem~\ref{theorem:basbm})

	Given a graph $G$ generated by a Binary Asymmetric SBM with two communities sized $\rho n$ and $(1-\rho)n$ for some constant $\rho$, and with $p = \frac{a\log{n}}{n}$ and $q = \frac{b\log{n}}{n}$, $\sqrt{a} - \sqrt{b(1+\log{\frac{a}{b}})} > \sqrt{c\log{\frac{a}{b}}/\epsilon}$, and $\tilde{h}(c/\epsilon) > 1$, $\M^{SDP}_{\Stbl}$ with $\delta = n^{-c}$ exactly recovers the ground-truth community $\sigma^*$, i.e., $\Pr[\M^{SDP}_{\Stbl}(G)\neq Y^*] = \nmo$.
\end{theorem}

\begin{corollary}
  \label{cor:threshold}
When $\rho = 1/2$, the condition for exact recovery is $\sqrt{a}-\sqrt{b}\geq \sqrt{2}\sqrt{1 + c/\epsilon}$ and  $\sqrt{a} - \sqrt{b(1+\log{\frac{a}{b}})} > \sqrt{c\log{\frac{a}{b}}/\epsilon}$.
\end{corollary}

\subsection{Censored Binary}

\textbf{Privacy model.}
The adjacency matrix $A(G)$ is defined as $A_{ij} = 0$ if there is no edge between $i$ and $j$. $A_{ij} = L_{ij}$ if there is an edge generated between $i$ and $j$.
In this section, we define the neighborhood between two graphs $G\sim G'$ if $A(G)$ and $A(G')$ differ by exactly two entries (due to the symmetry).
This privacy model can protect the existence (and the non-existence) of an arbitrary edge $(i,j)$, where any two neighboring adjacency matrices differ at element $ij$ (and $ji$): $A(G)_{ij} = 0$ (not an edge) and $A(G')_{ij} \neq 0$ (an edge).
It can also protect the label of an arbitrary edge $(i,j)$ whenever $(i,j)$ exists in the input graphs, that any two neighboring adjacency matrix differ as follows: $A(G)_{ij} = -1$ and $A(G')_{ij} = +1$.

\textbf{SDP Relaxation.}
We reuse the notation and arguments for the Binary Asymmetric model to form the following optimization problems for the Censored Binary model. Even though they do not look much different from the BASBM's SDP, they have vastly different characteristics, because the adjacency matrices $A$, in this case, implies much more topological information.

\begin{minipage}[t]{0.45\linewidth}%
  \vspace{-0.215in}
  \begin{align}
    \max_\sigma&\sum_{ij}A_{ij}\sigma_i\sigma_j \\
    \text{s.t. } &\sigma_i\in\{\pm 1\}, i\in[n]\nonumber
  \end{align}
\end{minipage}%
\hfill\vline\hfill
\begin{minipage}[t]{0.53\linewidth}%
  \vspace{-0.215in}
  \begin{align}
    \widehat{Y}_{SDP} =& \argmax_Y \langle A , Y \rangle \\
    \text{s.t. } &Y \succcurlyeq 0\nonumber\\
                       &Y_{ii} =1, i\in[n]\nonumber
  \end{align}
\end{minipage}%

\textbf{Assumptions of parameters.}
Let $p = a\log{n}/n$ for some fixed constant $a$ (in the random edge generation model $G(n,p)$). For the random label model: $h(\xi, a) = a(\sqrt{1-\xi} -\sqrt{\xi})^2 > 1$, or $h(\xi, a) = 1 + \Omega(1)$. 

\begin{definition}
  $G$ is called $\mathcal{C}$-concentrated if there exists a tuple $\C=(c_1, c_2)$ such that $G$ satisfies two conditions:

\begin{minipage}[t]{0.52\linewidth}%
  \begin{itemize}
  \item $\Vert A - \E[A]\Vert_2 \leq c_1 \sqrt{\log{n}}$
  \end{itemize}
\end{minipage}%
\begin{minipage}[t]{0.47\linewidth}%
  \begin{itemize}
  \item $d^*_i \geq c_2 \log{n}, i\in [n]$
  \end{itemize}
\end{minipage}%

where $d^*_i = \sum_{j = 1}^n \Aij\sigma^*_i\sigma^*_j$ for every $i \in [n]$.
\end{definition}

Lemma~\ref{lemma:cbsbm-concen-high-prob},~\ref{lemma:cbsbm-distance-logn}, and~\ref{lemma:cbsbm-optimal} are the analogues of Lemma~\ref{lemma:main-concen-high-prob},~\ref{lemma:main-concen-stable}, and~\ref{lemma:main-sdp-optimal} for BASBM. The first condition can be proved w.h.p. using the same strategy as in BASBM. The second condition, however, cannot be easily adapted. We can transform $d^*_i = \sum_{j=1}^{n-1}X_j$, where $X_j \overset{i.i.d.}{\sim} p(1-\xi)\beta_{+1} + p\xi\beta_{-1} + (1-p)\beta_0$, in which $\beta_x$ is the Dirac delta function at $x$. In this case, $d^*_i$ cannot be represented by the difference between two Binomial distributions as in BASBM, and we have to utilize Chernoff bound: $\Pr[\sum_{j=1}^{n}X_j < c_2\log{n}] \leq  \exp(-n\ell(\frac{c_2\log{n}}{n}))$ where the function $\ell(x)$ is defined as $\ell(x) = \sup_{\lambda \geq 0}-\lambda x - \log{\E[e^{-\lambda X}]}$. We then solve the supremum at $\lambda^*$ and substituting $x$ by $c_2\log{n}/n$, and utilizing the assumption $h(\xi,a) > 1$ to determine the tail probability, which is bounded by $n^{-1-\Omega(1)}$.

%
%

\begin{theorem}
  \label{threorem:main-cbsbm}

  (Complete proof in Theorem~\ref{threorem:cbsbm}) 
Given graph $G$ generated by a CBSBM as described above where $c/\epsilon < a$ and $h(\xi, a) > 1$, $M^{SDP}_{\Stbl}$ with $\delta = n^{-c}$ exactly recovers the ground-truth community $Y^*$, i.e., $\Pr[M^{SDP}_{\Stbl}(G)\neq Y^*] = \nmo$
\end{theorem}

\subsection{General Structure}

\textbf{SDP Relaxation.} Unlike the two other models, we use a set of indicator vectors $\xi_k$ to map a vertex to its cluster: $\xi_k(i) = 1$ if $i\in C_k$ and $\xi_k(i) = 0$ otherwise ($\xi_k$ refers to the variables while $\xi^*_k$ is the ground-truth). We form the following optimization problems: the MLE on the left and the transformed SDP Relaxation on the right.

\begin{minipage}[t]{0.48\linewidth}%
  \vspace{-0.215in}
  \begin{align}
    \max_\xi&\sum_{ij}A_{ij}\sum_{k\in[r]}\xi_k(i)\xi_k(j)\nonumber \\
    \text{s.t. } &\xi_k\in\{0,1\}^n, k\in[r]\nonumber\\
            &\xi_k^T\bm{1} = K_k, k\in[r]\nonumber\\
    &\xi^T_k\xi_{k'} = 0, k\neq k'
  \end{align}
\end{minipage}%
\hfill\vline\hfill
\begin{minipage}[t]{0.49\linewidth}%
  \vspace{-0.215in}
  \begin{align}
    \widehat{Z}_{SDP} =& \argmax_Y \langle A , Z \rangle \nonumber\\
    \text{s.t. } &Z \succcurlyeq 0\nonumber\\
                       &Z_{ii} \leq 1, i\in[n]\nonumber\\
                       &Z_{ij}\geq 0, i,j\in[n]\nonumber\\
   \langle I, Z \rangle &= \sum_{k\in[r]}K_i\nonumber\\
   \langle J, Z \rangle &= \sum_{k\in[r]}K_i^2
  \end{align}
\end{minipage}%

Due to the space limits, we only present the $\C$-concentration definition for GSSBM and the final exact recovery statement and its conditions (Theorem~\ref{theorem:main-gssbm}) here. We show the complete analysis in Section~\ref{sec:gssbm}.

\begin{definition}
  \label{def:gssbm-concen}
  $G$ is called $\C$-concentrated if there exists a tuple of constants $\C = (c_1, c_2, c_3, c_4, c_5)$ such that $G$ satisfies these conditions:
  \begin{itemize}
  \item $\Vert A(G) - \E[A(G)] \Vert \leq c_1\sqrt{\log{n}}$
  \item $\min_{i\in [n]}s_i \geq (b + 2c_2)\rho_{k(i)}\log{n}$
  \item $\max_{i\in[n], k: k \neq k(i)}e(i, C_{k}) \leq (b + c_2)K_{k}\log{n}/n - c_3\log{n}$
  \item $\min_{i,j: k(i)k(j)[k(i)-k(j)] \neq 0}e(C_{k(i)}, C_{k(j)})\geq K_{k(i)}K_{k(j)}q - 2\sqrt{K_{k(i)}K_{k(j)}}\sqrt{{\log{n}}} - c_4\log{n}$
  \item $\max_{i\in C_0}e(i, C_{k: k \neq 0}) \geq \tilde{\tau}K_r\frac{\log{n}}{n} - c_5\log{n}$,
  \end{itemize}
  where $k(i)$ is $i$'s cluster, $e(i, C_k)$ is the number of edges between a node $i$ and nodes from cluster $C_k$, $e(C_k,C_{k'})=\sum_{i\in C_k}e(i, C_{k'})$, $s_i = e(i,C_{k(i)})$, $\tilde{\tau} = b + 2c_2$.
\end{definition}



\begin{theorem}
  \label{theorem:main-gssbm}
  (Complete proof in Theorem~\ref{theorem:gssbm-exact-recovery})
Let $I(x,y) = x - y\log{\frac{ex}{y}}$.
Given a graph $G$ generated by a GSSBM as above, $M^{SDP}_{\Stbl}$ with $\delta = n^{-c}$ exactly recovers the ground-truth community $Z^*$, i.e., $\Pr[M^{SDP}_{\Stbl}(G)\neq Z^*] = \nmo$, if the following conditions are satisfied: $
      I(a, b+\frac{2c}{\epsilon\rho_{min}}) > 1/\rho_{min};
      I(b, b+ \frac{c}{\epsilon}(\frac{1}{\rho_{min}} - 1)) > 1/\rho_{min};
      I(b, b+ \frac{c}{\epsilon}(\frac{2}{\rho_{min}} - 1)) > 1/\rho_{min}.$
\end{theorem}

\section{Polynomial-time algorithm}
\label{sec:poly-alg}

In Algorithm~\ref{alg:stability-mechanism}, calculating $d_{SDP}(G)$ will takes at least $n^{O(\log{n})}$ times due to calculating $d(G)$. The main idea is if we can estimate $d(G)$ faster, we can design a faster algorithm. Observe that when $G$ is $\C$-concentrated, $d(G) \geq c\log{n}/\epsilon$. We test if the input graph is $\C$-concentrated and if the graph pass the test, we can set $\widehat{d}(G) = d(G) \geq c\log{n}/\epsilon$ and use $\widehat{d}$ instead of $d$. If $G$ fails the test, we compute $\widehat{d}(G) = \min(d(G), c\log{n}/\epsilon)$. The main challenge is that, testing $\C$-concentration requires knowledge of the SBMs, i.e., $p$, $q$ and \textbf{most importantly}, $\sigma^*$ (or $\xi^*_k$ in $r > 2$ communities)--the quantities we are trying to output. 

In several applications, the edge probabilities $p$ and $q$ (and $a, b$ respectively) may be known by the algorithms, which makes the problem easier.
When they are unknown, we need a reliable way to estimate them from the input.

Suppose that we have access to oracles that can provide us with these quantities. Let $\Oracle_{\sigma^*}$ be the one that can provide us the true value of $\sigma^*$ (or $\xi^*_k$ when $r>2$). Let $\Oracle_{a,b}^\alpha$ be the one that can provide us the parameters $\hat{a}, \hat{b}$ accurately up to a factor of $1\pm\alpha$ from the true values of $a, b$ for a small constant $\alpha < 0.001$. 

We present Algorithm~\ref{alg:oracle-mechanism} with the unrealistic assumption of the oracles. We then prove that Algorithm~\ref{alg:oracle-mechanism} ($\M_{\Stbl\Oracle}$) retains the privacy and utility of Algorithm~\ref{alg:stability-mechanism}. Because checking the $\C$-concentration can be done in polynomial time (in terms of $n$), and w.h.p.~we do not invoke $d(G)$, the Algorithm takes polynomial time w.h.p..~After we confirm that $\M_{\Stbl\Oracle}$ has all the properties we need, we replace the oracles by realistic alternatives that we calculate from the input graph $G$. We then prove that Algorithm~\ref{alg:poly-stability-mechanism} that w.h.p. is the same as Algorithm~\ref{alg:oracle-mechanism} and inherits all of its properties.


\begin{algorithm}
  \caption{$\M^{f}_{\Stbl \textsc{Fast}}(G, \C)$: Fast Stability Mechanism}
  \label{alg:poly-stability-mechanism}
  \begin{algorithmic}[1]
    \STATE $\hat{Y} \leftarrow f(G)$
    \STATE $\sigma^* \leftarrow \hat{Y}$
    \STATE $(\hat{a}, \hat{b}) \leftarrow Algorithm~\ref{alg:estimate}(G)$
    \STATE $\hat{\C} \leftarrow \text{ adjust }\C \text{ on }\alpha = 0.001 \text{ to satisfy Proposition~\ref{prop:oracle-concen}}$
    \STATE Construct $\hat{\C}$-concentration using $\hat{C}, \sigma^*, \hat{a}, \hat{b}$
    \IF {$G$ is $\hat{\C}$-concentrated}
    \STATE $\hat{d}(G) \leftarrow c\log{n}/\epsilon$
    \ELSE
    \STATE $\hat{d}(G) \leftarrow \min(c\log{n}/\epsilon, d(G))$
    \ENDIF
    \STATE $\tilde{d}(G) \leftarrow \hat{d}(G) + \Lap(1/\epsilon)$
    \IF{$\tilde{d}_f(G)  > \frac{\log{1/\delta}}{\epsilon}$}
    \STATE Output $\hat{Y}$
    \ELSE
    \STATE Output $\perp$ 
    \ENDIF
  \end{algorithmic}
\end{algorithm}

\begin{proposition}
  \label{prop:oracle-concen}
  In the context of Algorithm~\ref{alg:oracle-mechanism}, if a graph $G$ is $\hat{\C}$-concentrated then it is $\C$-concentrated.
\end{proposition}

 In each SBM setting, the proposition can be easily verified by checking all conditions of $\C$-concentration. $\Oracle_{\sigma^*}$~guarantees us the true value of $\sigma^*$, so the differences between $\C$ and $\hat{\C}$ only come from the factor $\alpha$ of $\Oracle_{a,b}^\alpha$. We use a tighter tuple of constants $\hat{\C}$ (compared to $\C$) to balance the fact that $\hat{a}$ and $\hat{b}$ may be off by some factor of $1\pm \alpha$. This task can be done by adjusting each condition by scaling the respective $c_k$ to a factor of $1\pm 2\alpha$ in which direction makes the condition tighter.

\begin{theorem}
\label{theorem:main-poly}  
(Complete proof in Lemma~\ref{lemma:oracle-privacy},~\ref{lemma:oracle-exact-recovery},~\ref{lemma:poly})
$\M_{\Stbl\textsc{Fast}}$ (Algorithm~\ref{alg:poly-stability-mechanism}) is $(\epsilon,\delta)$-DP. When $\M^{f}_{\Stbl}$ (Algorithm~\ref{alg:stability-mechanism}) achieves exact recovery (under some specific conditions of the SBMs under the view of Lemma~\ref{lemma:prelim-stability-mechanism}), $\M^f_{\Stbl\textsc{Fast}}$ also achieves exact recovery under the same conditions; and takes $O(poly(n))$ w.h.p.. 
\end{theorem}

\textbf{Proof sketch.} $\M_{\Stbl\Oracle}$ differs from $\M_{\Stbl\textsc{Fast}}$ by exact two steps: ($\sigma^* \leftarrow \Oracle_{\sigma^*}$ line 2)  and ($(\ahat,\bhat) \leftarrow \Oracle^{\alpha}_{a,b}$ line 3). We then complete three big steps to prove the Theorem. \textbf{First, } we prove that $\Delta_{\hat{d}} = 1$ (the global sensitivity of $\hat{d}$) and follows the arguments of Theorem~\ref{theorem:privacy}, substituting $d$ by $\hat{d}$ to confirm the privacy guarantee of $\M_{\Stbl\Oracle}$. \textbf{Second, } we utilizes Proposition~\ref{prop:oracle-concen}, arguing that $\hat{\C}$-concentration implies $\C$-concentration, to show that any graph $G$ passes the test at line 6 is $\log{n}$-stable. Since $\hat{\C}$-concentration is a valid concentration, $G$ passes the test at line 6 w.h.p.. In view of Lemma~\ref{lemma:prelim-stability-mechanism}, $\M_{\Stbl\Oracle}$ achieve exact recovery. \textbf{Third, } we replace the oracles (by line 2, 3 of Algorithm~\ref{alg:poly-stability-mechanism}), arguing that using $\hat{Y}$ is as good as $\Oracle_{\sigma^*}$ w.h.p.. Similarly, the estimator by Algorithm~\ref{alg:estimate} is at least as good as $\Oracle^\alpha_{\sigma^*}$ w.h.p. in view of Lemma~\ref{lemma:param-estimate}. We now can confirm that w.h.p., $\M_{\Stbl\textsc{Fast}}$ is as good as $\M_{\Stbl\Oracle}$ and complete the proof.

\section{Conclusion}

Our work studied the community detection problem in SBMs under differential privacy, focusing on the theoretical boundaries for exact recovery. We show that, in three variants of SBMs: Binary Asymmetric, Censored Binary, and General Structure, exact recovery is possible by composing the Stability mechanism and the Semi-definite programming estimator. The main challenges lie in the design and analysis of $\C$-concentration--a key concept that determines the stability and optimality of the SDP Relaxation. Our results extend the best theoretical boundaries for exact recovery in SBMs for symmetric variants. We also propose the first polynomial time algorithms for SBMs under edge DP with arbitrary small $\epsilon$ that matches the non-private's recovery threshold when $\epsilon\rightarrow\infty$.

\noindent
\textbf{Acknowledgements.}
This research is supported by University of Virginia Strategic Investment Fund award number SIF160, NSF Grants OAC-1916805 (CINES), CCF-1918656
(Expeditions), IIS-1931628, IIS-1955797, CNS-2317193 and NIH grant R01GM109718.


\clearpage

\section*{Impact Statement}

This paper presents work whose goal is to advance the field of Machine Learning. 
There are many potential societal consequences of our work, none which we feel must be specifically highlighted here.

\bibliographystyle{icml2024}
\bibliography{ref.bib}
\appendix

\section{Stability Mechanism}

\begin{theorem}
  \label{theorem:privacy}
  $\M^{f}_{\Stbl}$ is $(\epsilon, \delta)$-differentially private
\end{theorem}

\begin{proof} (Adapted from Lemma 3.2 of~\cite{pmlr-v162-mohamed22a} for completeness and consistency)

The proof that the stability based mechanism satisfies $(\epsilon, \delta))$- DP follows directly from \cite{dwork2014algorithmic}. Given a pair of neighbor graphs $G\sim G'$, $d(G)$ denotes the distance from $G$ to its nearest unstable instance and $d(G')$ is the distance from $G'$ to its nearest unstable instance. Due to the triangle inequality, $| d(G)-d(G')| \leq 1$, hence the sensitivity of $d$: $\Delta_d = 1$. Adding a Laplacian noise of magnitude of $1/\epsilon$ guarantees $\epsilon$-differential privacy for $\tilde{d}$. In order to verify $(\epsilon, \delta)$-DP for the overall mechanism, we consider two scenarios: the first one, when the output of the mechanism is $\perp$. In this case, we have:


 \begin{align}
    \Pr[\M^f_{\Stbl}(G) = \perp] &= \Pr \left[\tilde{d} (G) \leq  \frac{\log{1/\delta}}{\epsilon} \right] \\
                               &\leq e^\epsilon \Pr \left[\tilde{d}(G') \leq \frac{\log{1/\delta}}{\epsilon} \right] \\
    &=e^\epsilon \Pr[\M^f_{\Stbl}(G') = \perp].
  \end{align}
where the first inequality follows from the fact that $\tilde{d}(G)$ satisfies $\epsilon$-DP. For the second scenario, when the output of the mechanism $f(G)$, we have to analyze two cases. 

   The remaining part of the proof, we prove that output $f(G)$ in line 3 satisfies $(\epsilon, \delta)$-differential privacy to fulfill the proof of the theorem. We analyze two cases (1) $d(G) = 0$ and (2) $d(G) > 0$.

   \textbf{Case 1. $d(G)=0$}, we have $\Pr[\tilde{d}(G) > \frac{\log{1/\delta}}{\epsilon}] = \Pr[Lap(1/\epsilon) > \frac{\log{1/\delta}}{\epsilon}] \leq e^{-\log{1/\delta}} = \delta$. For any set of output $S \subseteq (Range(f) \cup \{\perp\})$, we have
   \begin{align*}
     \Pr[\M^f_{\Stbl}(G) \in S] &\leq \Pr[\M^f_{\Stbl}(G) \in (S\cup \{\perp\})] \\
     &\leq \Pr[\M^f_{\Stbl}(G) \in (S \cap \{\perp\})] + \Pr[\M^f_{\Stbl}(G) \neq \perp] \\
     &\leq \Pr[\M^f_{\Stbl}(G) \in (S \cap \{\perp\})] + \delta \\
     &\leq e^\epsilon\Pr[\M^f_{\Stbl}(G') \in (S \cap \{\perp\})] + \delta \\
     &\leq e^\epsilon\Pr[\M^f_{\Stbl}(G') \in S] + \delta, 
   \end{align*}
   where the third inequality is because $\Pr[\M^f_{\Stbl}(G) \neq \perp] = \Pr[\tilde{d}(G) > \frac{\log{1/\delta}}{\epsilon}] \leq \delta$ and the fourth inequality is because $S\cap \{\perp\}$ is \textbf{(a)} $\emptyset$ or \textbf{(b)} $\{\perp\}$. When $\textbf{(a)}$ happens, $\Pr[\\M^f_{\Stbl}(G) \in \emptyset] = \Pr[\M^f_{\Stbl}(G') \in \emptyset] = 0$ and when $\textbf{(b)}$ happens, it follows above proof that $\Pr[\M^f_{\Stbl}(G) = \perp] \leq e^\epsilon\Pr[\M^f_{\Stbl}(G') = \perp]$.
 
   \textbf{Case 2. $d(G)>0$}. In this case, $G$ is at least $1$-stable, which means: $\bm{\sigma}(G) = \bm{\sigma}(G') = \sigma$, we have:

   \begin{align*}
     \Pr[\M^f_{\Stbl}(G) = \sigma] &= \Pr[\tilde{d}(G) > \frac{\log{1/\delta}}{\epsilon}] \\
     &\leq e^\epsilon \Pr[\tilde{d}(G') > \frac{\log{1/\delta}}{\epsilon}] \\
                                     &= e^\epsilon\Pr[\M^f_{\Stbl}(G') = \sigma],
   \end{align*}

   and the Lemma follows.
\end{proof}

\begin{lemma}
  \label{lemma:stability-mechanism}
  (Full version of Lemma~\ref{lemma:prelim-stability-mechanism})
  Given a function $f: \mathcal{G} \rightarrow \mathcal{R}$, the $f$-based Stability mechanism with $\delta = n^{-t}$ has $\Pr[M^f(G) \neq f(G)] \leq n^{-k_1} + n^{-k_2}$, if a graph $G$ is $ \frac{t + k_1}{\epsilon}\log{n} $-stable under function $f$ with probability at least $1-n^{-k_2}$. When $k_1, k_2 = \Omega(1)$, $\Pr[M^f(G) \neq f(G)] \leq n^{-\Omega(1)}$.
\end{lemma}

\begin{proof}
  From the definition of the Stability mechanism, $M^f(G)$ does not output $f(G)$ when $\tilde{d}(G) \leq \frac{\log{1/\delta}}{\epsilon.}$. For simplicity, we denote $\tilde{d}(G)$ and $d(G)$ as $\tilde{d}, d$ respectively in the following equations. Hence, we have:

  \begin{align}
    \Pr[M^f(G) \neq f(G)] &= \Pr[\tilde{d} < \frac{\log{1/\delta}}{\epsilon}]\\
                          &= \Pr[d + \Lap(1/\epsilon) < \frac{\log{1/\delta}}{\epsilon}]\\
                          &= \Pr[\Lap(1/\epsilon) < \frac{\log{1/\delta}}{\epsilon}- d]\\
                          &= \Pr[\Lap(1/\epsilon) < \frac{\log{1/\delta}}{\epsilon}- d | d \geq \frac{t + k_1}{\epsilon}\log{n}]\Pr[d \geq \frac{t + k_1}{\epsilon}\log{n}] \\&+ \Pr[\Lap(1/\epsilon) < \frac{\log{1/\delta}}{\epsilon}- d | d < \frac{t + k_1}{}\log{n}]\Pr[d < \frac{t + k_1}{\epsilon}\log{n}]\\
    &\leq \Pr[\Lap(1/\epsilon) < \frac{\log{1/\delta}}{\epsilon}- d | d \geq \frac{t + k_1}{\epsilon}\log{n}] + \Pr[d < \frac{t + k_1}{\epsilon}\log{n}]\\
    &\leq \Pr[\Lap(1/\epsilon) < \frac{\log{1/\delta}}{\epsilon}- d | d \geq \frac{t + k_1}{\epsilon}\log{n}] + n^{-k_2}\\
    &\leq \Pr[\Lap(1/\epsilon) < \frac{t\log{n}- (t + k_1)\log{n}}{\epsilon}] + n^{-k_2}\\
    &\leq \Pr[\Lap(1/\epsilon) < \frac{-k_1\log{n}}{\epsilon}] + n^{-k_2}\\
    &= \Pr[\Lap(1/\epsilon) < \frac{-k_1\log{n}}{\epsilon}] + n^{-k_2}\\
    &\leq n^{-k_2} + n^{-k_2},.
  \end{align}
  and the Lemma follows.
\end{proof}


\section{Binary Asymmetric SBM (BASBM)}


In this section we examine the Binary Asymmetric SBM (referred to as BASBM) where the sizes of the clusters are defined by a parameter $\rho \in (0, 1/2)$. We assume that the first cluster has size $K = \rho n$ and the second cluster has size $n - K = (1-\rho) n$.

\textbf{Problem Formulation and Definitions}
We adapt~\cite{hajek2016extensions}'s analyses to formulate the Semi-definite program to solve the community detection in the BASBM.

\begin{definition}
	\label{def:binary}
	We define quantities in our analysis as follows:
	\begin{itemize}
		\item $Y = \sigma\sigma^T$
		\item $Y^* = \sigma^*\sigma^{*T}$
		\item $\tau = \frac{a-b}{\log{a}-\log{b}}$
		\item $\lambda^* = \tau\log{n}/n$
		\item $d^*\in\mathbb{R}^n: d^*_i = \sum_{j=1}^n\Aij\sigma^*_i\sigma^*_j - \lambda^*(2K - n)\sigma^*_i$
		\item $D^* = diag\{d^*\}$
		\item $I$ is the identity matrix
		\item $J$ is the all-one matrix
		\item $\J(x) = \left(\lambda^* - \frac{p+q}{2}\right)x^\T Jx$
		\item $\xcheck = \arg \max_{x}x^\T Jx$ subject to $\Vert x \Vert_2 = 1$ and $\langle x, \sigma^* \rangle = 0$
		\item $h(\alpha) = a\rho + b(1-\rho) - \sqrt{\alpha^2 + 4\rho(1-\rho)ab} + \frac{|\alpha|}{2}\log{\frac{\rho b}{(1-\rho)a}}$
		\item $\tilde{h}(x) = a\rho + b(1-\rho) - \sqrt{(\tau(1-2\rho) - x)^2 + 4\rho(1-\rho)ab} + \frac{\tau(1-2\rho) - x}{2}\log{\frac{\rho b}{(1-\rho)a}} $
	\end{itemize}
\end{definition}

\begin{definition}
	\textbf{Definition of \emph{Concentration}.}

	We assume a graph $G$ is generated by a Binary Symmetric SBM with two communities sized $\rho n$ and $(1-\rho)n$ for some constant $\rho$. The link between two endpoints from the same community are generated with probability $p = \frac{a\log{n}}{n}$ and the link between two endpoints from different communities are generated with probability $q = \frac{b\log{n}}{n}$.

	$G$ is called $(c_1, c_2, c_3, c_4)$-concentrated for some constants $c_i > 0$ if $G$ satisfies all $4$ conditions:

	\begin{itemize}
		\item $\Vert A - \E[A]\Vert_2 \leq c_1\sqrt{\log{n}}$
		\item $\xcheck^\T D^*\xcheck + \J(\xcheck) > c_2\log{n} $
		\item $\Vert (D^* - \E[D^*])\xcheck\Vert_2 \leq c_3\sqrt{\log{n}}$
		\item $d^*_i \geq c_4\log{n}$ for every $i\in[n]$
	\end{itemize}
\end{definition}

\textbf{Assumptions of the parameters}

We have several assumptions of the parameters of the SBM. These assumptions will be used in our analyses to derive the conditions in which Exact Recovery under Differential Privacy is feasible.

The exists a constant $c > 0$ such that
\begin{align}
	\tilde{h}(c) > 1 \text{ or equivalently, } h(c -\tau(1-2\rho)) > 1
	\label{eq:assumption-h}
\end{align}

In other words, we can say that:

\begin{align}
	\tilde{h}(c) = 1 + \Omega(1) \text{ or equivalently, } h(c -\tau(1-2\rho)) = 1 + \Omega(1)
	\label{eq:assumption-h}
\end{align}

The Binary Asymmetric SBM can be solve in the non-privacy setting by solving the following SDP relaxation. We denote $SDP(G)$ as a function taking input graph $G$ and outputting the optimal solution of the SDP relaxation constructed by its adjacency matrix $A(G)$.

\begin{definition} SDP Relaxation of the Binary Asymmetric SBM:

\end{definition}
\begin{align}
  \widehat{Y}_{SDP} &= \argmax_Y \langle A, Y \rangle \\
  \text{s.t.  }Y &\curlyeqsucc 0 \\
  Y_{ii} &= 1, \text{ for } i \in [n] \\
  \langle J, Y \rangle &= (2K-n)^2
\end{align}

\begin{lemma}
	\label{lemma:adj-mat-concen}
	\textbf{Theorem 5 of~\cite{hajek2016achieving}}. Let $A$ be a symmetric and zero-diagonal random matrix, where the entries $\Aij (i<j)$ are independent and $[0, 1]$-valued. Assume that $\E[\Aij] \leq p$, where $c_0\log{n}/n \leq p \leq 1-c_1$ for arbitrary constants $c_0,c_1 > 0$. Then for any $c> 0$, there exists $c'>0$ such that for any $n\geq 1, \Pr[\Vert A - \E[A]\Vert_2 \leq c'\sqrt{np}] \geq 1-n^{-c}$.
\end{lemma}

\begin{lemma}
	\label{lemma:talagrand}
	\textbf{Theorem 2.1.13 of~\cite{tao2011topics}} (Originally from~\cite{talagrand1995concentration}). Let $P>0$, and let $X_1, \ldots, X_n$ be independent complex variables with $|X_i| \leq P$ for all $1 \leq i \leq n$. Let $F: \mathbf{C}^n \rightarrow \mathbf{R}$ be a $1$-Lipschitz convex function. Then for any $\lambda$ one has
	\begin{align}
		\Pr[|F(X) - \E[F(X)]| \geq \lambda P] \leq C \exp(-c\lambda^2)
	\end{align}
	for some absolute constants $C, c >0$.
\end{lemma}

\begin{lemma}
	\label{lemma:binom-diff}
	\textbf{Lemma 2 of~\cite{hajek2016extensions}}. Suppose $a, b > 0, \alpha \in \mathbb{R}$, and $\rho_1, \rho_2 > 0$. Let $X, R$ be independent with $X \sim Binom(m_1, \frac{a\log{n}}{n})$ and $R\sim Binom(m_2, b)$, where $m_1=\rho_1n + o(n)$ and $m_2 = \rho_2n + o(n)$ as $n\rightarrow\infty$. Let $k\in\mathbb{I}$ such that $k = \alpha\log{n} + o(\log{n})$. If $\alpha \leq a\rho_1 - b\rho_2$,

	\begin{align}
		\Pr[X - R \leq k] = n^{-g(\rho_1, \rho_2, a, b, \alpha) + o(1)},
	\end{align}

	where $g(\rho_1, \rho_2, a, b, alpha) = a\rho_1 + b\rho_2 -\gamma - \frac{\alpha}{2}\log{\frac{(\gamma-\alpha)a\rho_1}{(\gamma+\alpha)b\rho_2}}$ with $\gamma = \sqrt{\alpha^2 + 4\rho_1\rho_2ab}$.

	Furthermore, for any $m_1, m_2, \in \mathbb{N}$, $k \in \mathbb{I}$ such that $k < (m_1a - m_2b)\log{n}/n$,

	\begin{align}
		\Pr[X-R \leq k] \leq n^{-g(m_1/n, m_2/n, a, b, k/\log{n})}
	\end{align}
\end{lemma}

\begin{lemma}
	\label{lemma:second-cond}
	A graph $G$ generated by a Binary Asymmetric SBM with two communities sized $\rho n$ and $(1-\rho)n$ for some constant $\rho$ and with $p = \frac{a\log{n}}{n}$ and $q = \frac{b\log{n}}{n}$, there exists some constant $c_2: 0 <c_2 < \tau - b$ such that
	\begin{align}
		\xcheck^\T D^*\xcheck + \J(\xcheck) > c_2\log{n},
	\end{align}

	with probability at least $1-\nmo$.
\end{lemma}

\begin{proof}

	\textbf{The second condition} relies on Talagrand's concentration inequality for Lipschitz convex functions (Lemma~\ref{lemma:talagrand}). For the context of Lemma~\ref{lemma:talagrand}, set the function $F(A) = \xcheck^\T D^*\xcheck$ where the adjacency matrix $A$ is the argument of the function $F$ (as $D^*$ can be represented as a mapping of $A$). The next step is to prove that $F(A)$ is Lipschitz continuous in $A$.

	We note that $\xcheck^\T D^*\xcheck = \langle A, B \rangle - \lambda^*(2K-n)\sum_{i=1}^n\xcheck_i^2\sigma^*_i$ where $B_{ij} = \sigma^*_i\sigma^*_j\xcheck_i^2$, which makes $\xcheck^\T D^*\xcheck$ is Lipchitz continuous with Lipschitz constant $\Vert B \Vert_F= \sqrt{\frac{(1-\rho)^2}{\rho} + \frac{\rho^2}{1-\rho}} + o(1)$
    We treat all $A_{ij}$ as $X$ in Lemma~\ref{lemma:talagrand} with $P = 1$ (as $0 \leq \Aij \leq 1$). By Lemma~\ref{lemma:talagrand}, for any $c$ there exists some constants $C, c_2'$ and $\lambda = c_2'\sqrt{\log{n}}$ that:

	\begin{align}
		\Pr[|\xcheck^\T D^*\xcheck - \E[\xcheck^\T D^*\xcheck]| \geq c_2'\sqrt{\log{n}}] \leq C\exp(-c\log{n})
	\end{align}

	It follows that if we pick any constant $c > 0$, then with probability at least $1-\nmo$, we have:
	\begin{align}
		\xcheck^\T D^*\xcheck - \E[\xcheck^\T D^*\xcheck] > -c_2'\sqrt{\log{n}}
		\label{eq:xcheck-concen}
	\end{align}

	We now analyze $\J(\xcheck)$. Because $\xcheck^\T J\xcheck = 4K(n-K)/n$, we have:

	\begin{align}
		\J(\xcheck) & = \left(\lambda^* - \frac{p+q}{2}\right)\xcheck^T J\xcheck \\
		            & = \left(\tau - \frac{a+b}{2}\right)4K(n-K)\log{n}/n^2      \\
		            & =(2\tau - a - b)2K(n-K)\log{n}/n^2                         \\
		            & =(\tau - a)2K(n-K)\log{n}/n^2 + (\tau-b)2K(n-K)\log{n}/n^2
		\label{eq:xcheck-j}
	\end{align}

	And then we analyze $\E[\xcheck^\T D^*\xcheck]$:

	In the first case: $\sigma^*_i = 1$:

	\begin{align}
		\E[d^*_i] \overset{\text{def}}{=} \bar{d}_+ = \left( K(a-\tau) + (n-K)(\tau - b) -a \right)\log{n}/n
	\end{align}

	In the second case: $\sigma^*_i = -1$:

	\begin{align}
		\E[d^*_i] \overset{\text{def}}{=} \bar{d}_- = \left( (n-K)(a-\tau) + (K(\tau - b) -a \right)\log{n}/n
	\end{align}

	Then we have:

	\begin{align}
		\E[\xcheck^\T D^*\xcheck] & = (n-K)\bar{d}_+/n + K\bar{d}_-/n                                                                \\
		                          & = \left( 2K(n-K)(a-\tau) + (K^2 + (n-K)^2)(\tau - b) - na \right)\log{n}/n^2                     \\
		                          & =2K(n-K)(a-\tau)\log{n}/n^2+ K^2(\tau -b)\log{n}/n^2 + (n-K)^2(\tau - b)\log{n}/n^2 - a\log{n}/n
		\label{eq:xcheck-expect}
	\end{align}.

	Compose with the result of equation~\ref{eq:xcheck-j}, taking the sum of the two quantities, we have:

	\begin{align}
		\E[\xcheck^\T D^*\xcheck] + \J(\xcheck) & = \frac{(\tau - b)\log{n}}{n^2} \left( K^2 + (n-K)^2 +2K(n-K) \right) - \frac{a\log{n}}{n} \\
		                                        & =(\tau-b)\log{n} - \frac{a\log{n}}{n}
		\label{eq:xcheck-expect-j}
	\end{align}

	Back to the second condition, composing the results of equations~\ref{eq:xcheck-concen},~\ref{eq:xcheck-expect},~\ref{eq:xcheck-j},~\ref{eq:xcheck-expect-j}, we rewrite the second condition as follows:

	\begin{align}
		\xcheck^\T D^*\xcheck + \J(\xcheck) & = \xcheck^\T D^*\xcheck - \E[\xcheck^\T D^*\xcheck ] + E[\xcheck^\T D^*\xcheck ] + \J(\xcheck) \\
		                                    & \geq -c_2'\sqrt{\log{n}} + (\tau-b)\log{n} - \frac{a\log{n}}{n}                                \\
		                                    & > c_2\log{n},
	\end{align}

	for some constants $0 < c_2 < \tau - b$ and $n$ large enough. Since equation~\ref{eq:xcheck-concen} holds with probability at least $1-\nmo$, the Lemma follows.
\end{proof}

\begin{lemma}
	\label{lemma:third-cond}
	A graph $G$ generated by a Binary Asymmetric SBM with two communities sized $\rho n$ and $(1-\rho)n$ for some constant $\rho$ and with $p = \frac{a\log{n}}{n}$ and $q = \frac{b\log{n}}{n}$, there exists some tuples of constant $c_3$ such that
	\begin{align}
		\Vert (D^* - \E[D^*]\Vert_2)\xcheck \leq c_3\sqrt{\log{n}},
	\end{align}

	with probability at least $1-\nmo$.
\end{lemma}

\begin{proof}
	\textbf{The third condition} can be constructed in a similar way to the second condition. First, we have:

	\begin{align}
		\Vert (D^* - \E[D^*])\xcheck \Vert_2^2 & = \sum_i\left(\sum_j(\Aij - \E[\Aij])\sigma^*_i\sigma^*_j\xcheck-i\right)^2       \\
		                                       & = \sum_i\xcheck_i^2{\sigma^*_i}^2\left(\sum_j(\Aij - \E[\Aij])\sigma^*_j\right)^2 \\
		                                       & = \sum_i\xcheck_i^2\left(\sum_j(\Aij - \E[\Aij])\sigma^*_j\right)^2
	\end{align}

	Then we analyze the expectation of $\Vert (D^* - \E[D^*])\xcheck \Vert_2$:

	\begin{align}
		\E[\Vert(D^* - \E[D^*])\xcheck \Vert_2] & \leq \sqrt{\E\left[\Vert(D^* - \E[D^*])\xcheck \Vert_2^2\right]}                          \\
		                                        & = \sqrt{\E\left[\sum_i\xcheck_i^2\left(\sum_j(\Aij - \E[\Aij])\sigma^*_j\right)^2\right]} \\
		                                        & =\sqrt{\sum_i\xcheck_i^2\E\left[\left(\sum_j(\Aij - \E[\Aij])\sigma^*_j\right)^2\right]}  \\
		                                        & \overset{(a)}{=}\sqrt{\sum_i\xcheck_i^2\sum_j\text{Var}(\Aij)}                            \\
		                                        & \leq \sqrt{\sum_i\xcheck_i^2\sum_j\frac{a\log{n}}{n}\left(1-\frac{a\log{n}}{n}\right)}    \\
		                                        & \leq \sqrt{\sum_i\xcheck_i^2a\log{n}}                                                     \\
		                                        & =\sqrt{a\log{n}\sum_i\xcheck_i^2}                                                         \\
		                                        & =\sqrt{a\log{n}},
	\end{align}

	where $(a)$ is because of:

	\begin{align}
		                 & \E\left[\left(\sum_j(\Aij - \E[\Aij])\sigma^*_j\right)^2\right]                                                             \\
		=                & \left(\E\left[\sum_j(\Aij - \E[\Aij])\sigma^*_j\right]\right)^2 + \text{Var}\left(\sum_j(\Aij - \E[\Aij])\sigma^*_j \right) \\
		\overset{(b)}{=} & \left(\sum_j\E[(\Aij - \E[\Aij])]\sigma^*_j\right)^2 + \sum_j\text{Var}\left((\Aij - \E[\Aij])\sigma^*_j \right)            \\
		=                & 0 + \sum_j{\sigma^*_j}^2\text{Var}(\Aij - \E[\Aij])                                                                         \\
		=                & \sum_j\text{Var}(\Aij),
	\end{align}

	where $(b)$ is because of linearity of expectations and $\Aij$s are independent.

	Next we prove that $\Vert(D^*-\E[D^*])\xcheck \Vert_2$ is convex and Lipschitz continuous in A with Lipschitz constant bounded by $\max\left\{\sqrt{\frac{1-\rho}{\rho}}, \sqrt{\frac{\rho}{1-\rho}}\right\}$. For any $A, A'$, let $D^*, {D^*}'$ be the respectively diagonal matrices, we have:

	\begin{align}
		\left| \Vert(D^*-\E[D^*]\xcheck) \Vert_2 - \Vert({D^*}'-\E[{D^*}']\xcheck) \Vert\_2 \right|
		 & \leq \Vert (D^* - {D^*}')\xcheck \Vert_2                                                                                         \\
		 & =\sqrt{\sum_i{\xcheck_i^2\left(\sum_j(\Aij-\Aij')\sigma^*_j)\right)^2}}                                                          \\
		 & \leq\sqrt{\sum_i\left(\sum_j(\Aij-\Aij')\sigma^*_j \right)^2}\max\left\{ \sqrt{\frac{n-K}{nK}}, \sqrt{\frac{K}{n(n-K)}} \right\} \\
		 & \leq\Vert A - A'\Vert_F\max\left\{ \sqrt{\frac{1-\rho}{\rho}}, \sqrt{\frac{\rho}{1-\rho}} \right\}
	\end{align}

	Then we apply Lemma~\ref{lemma:talagrand}, with $P = 1$, for any $c > 0$ there exists $C, c_3' > 0$ and $\lambda = c_3'\sqrt{\log{n}}$ such that:

	\begin{align}
		\Pr[|\Vert(D^*-\E[D^*])\xcheck \Vert_2 - \E[\Vert(D^*-\E[D^*])\xcheck \Vert_2]| \geq c_3'\sqrt{\log{n}}] \leq C\exp(c\log{n})
	\end{align}

	Choosing a constant $c > 0$, with probability at least $1-\nmo$, we have:

	\begin{align}
		\Vert(D^*-\E[D^*])\xcheck \Vert_2 - \E[\Vert(D^*-\E[D^*])\xcheck \Vert_2 < c_3'\sqrt{\log{n}}
	\end{align}

	As before we prove that $\E[\Vert(D^*-\E[D^*])\xcheck \Vert_2 \leq \sqrt{a\log{n}}$, setting $c_3 = c_3' + \sqrt{a}$, we have that with probability at least $1-\nmo$:

	\begin{align}
		\Vert(D^*-\E[D^*])\xcheck \Vert_2 & \leq c_3\sqrt{\log{n}}
		\label{eq:third-concen}
	\end{align}
\end{proof}

\begin{lemma}
	\label{lemma:h}
	Let function $h$ and $g$ be defined as above, with fixed values of $\tau, \rho, a,b$ and some constants $c_4$, we have:

	\begin{align}
		h(-\tau(1-2\rho)+c4) & < g(\rho, 1-\rho, a, b, -\tau(1-2\rho)+c_4) \text{ ,and} \\
		h(\tau(1-2\rho)+c4)  & < g(1-\rho, \rho, a, b, -\tau(1-2\rho)+c_4).
	\end{align}
	Furthermore, for function $h(\alpha)$, if $|\alpha_1| < |\alpha_2|$ then $h(\alpha_1) > h (\alpha_2)$.
\end{lemma}

\begin{proof}
	We prove the first inequality as follows: setting $\gamma = \sqrt{(-\tau(1-2\rho)+c_4)^2 + 4\rho(1-\rho)ab}$ the definition of $g$, if $c4 < \tau(1-2\rho)$ we have:

	\begin{align}
		     & g(\rho, 1-\rho, a, b, -\tau(1-2/\rho) + c_4)                                                                                                         \\
		=    & a\rho + b(1-\rho) -\gamma - \frac{-\tau(1-2\rho)+c_4}{2}\log{ \frac{(\gamma - (-\tau(1-2\rho)+c_4))a\rho}{(\gamma + (-\tau(1-2\rho)+c_4))b(1-\rho)}} \\
		=    & a\rho + b(1-\rho) -\gamma + \frac{\tau(1-2\rho)-c_4}{2}\log{ \frac{(\gamma + (\tau(1-2\rho)-c_4))a\rho}{(\gamma - (\tau(1-2\rho)-c_4))b(1-\rho)}}    \\
		\geq & a\rho + b(1-\rho) -\gamma + \frac{\tau(1-2\rho)-c_4}{2}\log{ \frac{a\rho}{b(1-\rho)}}, \text{ since } \tau(1-2\rho)-c_4 \geq 0                       \\
		\geq & a\rho + b(1-\rho) -\gamma + \frac{\tau(1-2\rho)-c_4}{2}\log{ \frac{b\rho}{a(1-\rho)}}, \text{ since } a > b                                          \\
		=    & a\rho + b(1-\rho) -\sqrt{(\tau(1-2\rho)-c_4)^2 + 4\rho(1-\rho)ab} + \frac{\tau(1-2\rho)-c_4}{2}\log{ \frac{b\rho}{a(1-\rho)}}                        \\
		=    & h(-\tau(1-2\rho) + c_4).
	\end{align}

    In case $c_4 > \tau(1-2\rho)$, we have:
	\begin{align}
		     & g(\rho, 1-\rho, a, b, -\tau(1-2/\rho) + c_4)                                                                                                         \\
		=    & a\rho + b(1-\rho) -\gamma - \frac{-\tau(1-2\rho)+c_4}{2}\log{ \frac{(\gamma - (-\tau(1-2\rho)+c_4))a\rho}{(\gamma + (-\tau(1-2\rho)+c_4))b(1-\rho)}} \\
		=    & a\rho + b(1-\rho) -\gamma + \frac{\tau(1-2\rho)-c_4}{2}\log{ \frac{(\gamma + (\tau(1-2\rho)-c_4))a\rho}{(\gamma - (\tau(1-2\rho)-c_4))b(1-\rho)}}    \\
		\geq & a\rho + b(1-\rho) -\gamma + \frac{\tau(1-2\rho)-c_4}{2}\log{ \frac{a\rho}{b(1-\rho)}}, \text{ since } \tau(1-2\rho)-c_4 \leq 0                       \\
		\geq & a\rho + b(1-\rho) -\gamma + \frac{\tau(1-2\rho)-c_4}{2}\log{ \frac{a(1-\rho)}{b\rho}}, \text{ since } 1-\rho > \rho                                          \\
		= & a\rho + b(1-\rho) -\gamma + \frac{c_4 - \tau(1-2\rho)}{2}\log{\frac{b\rho}{a(1-\rho)}}\\
		=    & a\rho + b(1-\rho) -\sqrt{(\tau(1-2\rho)-c_4)^2 + 4\rho(1-\rho)ab} + \frac{c_4 - \tau(1-2\rho)}{2}\log{ \frac{b\rho}{a(1-\rho)}}                        \\
		=    & h(-\tau(1-2\rho) + c_4).
	\end{align}

	Similarly, the second inequality can be proved as follows: setting $\gamma = \sqrt{(\tau(1-2\rho)+c_4)^2 + 4\rho(1-\rho)ab}$ (note that this $\gamma$ is different from the one in the first inequality above), we have:

	\begin{align}
		     & g(1-\rho, \rho, a, b, \tau(1-2/\rho) + c_4)                                                                                                       \\
		=    & a(1-\rho) + b\rho -\gamma - \frac{\tau(1-2\rho)+c_4}{2}\log{ \frac{(\gamma - (\tau(1-2\rho)+c_4))a(1-\rho)}{(\gamma + (\tau(1-2\rho)+c_4))b\rho}} \\
		=    & a(1-\rho) + b\rho -\gamma + \frac{\tau(1-2\rho)+c_4}{2}\log{ \frac{(\gamma + (\tau(1-2\rho)+c_4))b\rho}{(\gamma - (\tau(1-2\rho)+c_4))a(1-\rho)}} \\
		\geq & a(1-\rho) + b\rho -\gamma + \frac{\tau(1-2\rho)+c_4}{2}\log{ \frac{b\rho}{a(1-\rho)}}, \text{ since } \tau(1-2\rho)+c_4 \geq 0                    \\
		\geq & a\rho + b(1-\rho) -\gamma + \frac{\tau(1-2\rho)+c_4}{2}\log{ \frac{b\rho}{a(1-\rho)}}, \text{ since } a > b \text{ and } \rho \leq 1-\rho         \\
		=    & h(\tau(1-2\rho) + c_4).
	\end{align}

	Finally, for $|\alpha_1| < |\alpha_2|$, we have:
	\begin{align}
		h(\alpha_1) - h(\alpha_2) & = -\sqrt{\alpha_1^2 + 4\rho(1-\rho)ab} + \frac{|\alpha_1|}{2}\log{ \frac{b\rho}{a(1-\rho)}} + \sqrt{\alpha_2^2 + 4\rho(1-\rho)ab} - \frac{|\alpha_2|}{2}\log{ \frac{b\rho}{a(1-\rho)}} \\
		                          & > 0 + \frac{\log{ \frac{b\rho}{a(1-\rho)}}}{2}(|\alpha_1| - |\alpha_2|)                                                                                                                \\
		                          & \geq 0,
	\end{align}
	where the last inequality is because $\frac{b\rho}{a(1-\rho)} < 1$ then $\log{ \frac{b\rho}{a(1-\rho)}} < 0$ and $|\alpha_1| - |\alpha_2| < 0$.

\end{proof}

%
%

\begin{lemma}
	\label{lemma:fourth-cond}
	A graph $G$ generated by a Binary Asymmetric SBM with two communities sized $\rho n$ and $(1-\rho)n$ for some constant $\rho$ and with $p = \frac{a\log{n}}{n}$ and $q = \frac{b\log{n}}{n}$, there exists some tuples of constant $c_4$ such that
	\begin{align}
		\min_{i\in[n]}d^*_i \geq c_4\log{n},
	\end{align}

	with probability at least $1-\nmo$.
\end{lemma}

\begin{proof}

	\textbf{The fourth condition} will be analyzed in two cases: for nodes in the first cluster ($\sigma^*_i = 1$) and for nodes in the second cluster ($\sigma^*_i = -i$).

	\textbf{The first case: $\sigma^*_i = 1$}. We first set two random variables $X \sim Binom(K-1, \frac{a\log{n}}{n})$ and $R \sim Binom(n-K, \frac{b\log{n}}{n})$. Setting $m_1 = K-1$, $m_2 = n-K$, $k = -\tau(1-2\rho)\log{n} + c_4\log{n}, \alpha = -(\tau(1-2\rho) + c_4)$. Fix a node $i$ in the first cluster, we have $\sum_j\Aij\sigma^*_i\sigma^*_j = X - R$. Applying Lemma~\ref{lemma:binom-diff} and selecting some constants $c_4 > 0)$ that satisfies Equation~\ref{eq:assumption-h}, we have:

	\begin{align}
		\Pr\left[\sum_j\Aij\sigma^*_i\sigma^*_j \leq -\tau(1-2\rho)\log{n} + c_4\log{n}\right] & \leq n^{-g(\rho,1-\rho, a, b, -\tau(1-2\rho) + c_4)} \\
		                                                                                       & \overset{(a)}{\leq} n^{-h(-\tau(1-2\rho) + c_4)}     \\
		                                                                                       & \overset{(b)}{=}n^{-\tilde{h}(c_4)}                  \\
		                                                                                       & \overset{(c)}{\leq} n^{-1 - \Omega(1)},
	\end{align}

	where $(a)$ is because of the result Lemma~\ref{lemma:h} such that $h(-\tau(1-2\rho) + c_4) \leq g(\rho,1-\rho, a, b, -\tau(1-2\rho) + c_4)$, and $(b)$ is because of the definitions of $h$ and $\tilde{h}$, and $(c)$ is the assumption at equation~(\ref{eq:assumption-h}).

	With the assumption that $\tilde{h}(c_4) > 1$, applying union bound on all nodes $i$ in the first cluster we have that with probability at least $1-\nmo$:

	\begin{align}
		\sum_j\Aij\sigma^*_i\sigma^*_j > -\tau(1-2\rho)\log{n} + c_4\log{n}
	\end{align}

	Recall that $d^*_i = \sum_j\Aij\sigma^*_i\sigma^*_j - \sigma^*_i\tau(2K-n)\log{n}/n$, with $\sigma^*_i = 1$, we have:

	\begin{align}
		d^*_i & = \sum_j\Aij\sigma^*_i\sigma^*_j + \tau(1-2\rho)\log{n}     \\
		      & > -\tau(1-2\rho)\log{n} + c_4\log{n} + \tau(1-2\rho)\log{n} \\
		      & = c_4\log{n}.
		\label{eq:first-cluster}
	\end{align}

	\textbf{The second case: $\sigma^*_i = -1$}. We set two random variables $X \sim Binom(n-K-1, \frac{a\log{n}}{n})$ and $R \sim Binom(K, \frac{b\log{n}}{n})$. Setting $m_1 = n-K-1$, $m_2 = K$, $k = \tau(1-2\rho)\log{n} + c_4\log{n}, \alpha = \tau(1-2\rho) + c_4$. Fix a node $i$ in the second cluster, we have $\sum_j\Aij\sigma^*_i\sigma^*_j = X - R$. Applying Lemma~\ref{lemma:binom-diff}, we have:

	\begin{align}
		\Pr\left[\sum_j\Aij\sigma^*_i\sigma^*_j \leq \tau(1-2\rho)\log{n} + c_4\log{n}\right] & \leq n^{-g(1-\rho, \rho, a, b, \tau(1-2\rho) + c_4)} \\
		                                                                                      & \overset{(a)}{\leq} n^{-h(\tau(1-2\rho) + c_4)}      \\
		                                                                                      & \overset{(b)}{\leq} n^{-h(-\tau(1-2\rho) + c_4)}     \\
		                                                                                      & \overset{(c)}{=} n^{-\tilde{h}(c_4)}                 \\
		                                                                                      & \overset{(d)}{\leq} n^{-1-\Omega(1)},
	\end{align}

	where $(a)$ is because of the result Lemma~\ref{lemma:h} such that $h(\tau(1-2\rho) + c_4) \leq g(1-\rho,\rho, a, b, \tau(1-2\rho) + c_4)$, and $(b)$ is because of the result of Lemma~\ref{lemma:h} such that such that $h(-\tau(1-2\rho) + c_4) \leq h(\tau(1-2\rho) + c_4)$, and $(c)$ is because of the definitions of $h$ and $\tilde{h}$, and $(d)$ is the assumption at equation~(\ref{eq:assumption-h}).

	With the assumption that $\tilde{h}(c_4) > 1$, applying union bound on all nodes $i$ in the second cluster we have that with probability at least $1-\nmo$:

	\begin{align}
		\sum_j\Aij\sigma^*_i\sigma^*_j > \tau(1-2\rho)\log{n} + c_4\log{n}
	\end{align}

	Recall that $d^*_i = \sum_j\Aij\sigma^*_i\sigma^*_j - \sigma^*_i\tau(2K-n)\log{n}/n$, with $\sigma^*_i = -1$, we have:

	\begin{align}
		d^*_i & = \sum_j\Aij\sigma^*_i\sigma^*_j - \tau(1-2\rho)\log{n}    \\
		      & > \tau(1-2\rho)\log{n} + c_4\log{n} - \tau(1-2\rho)\log{n} \\
		      & = c_4\log{n}.
		\label{eq:second-cluster}
	\end{align}

	Composing equations~\ref{eq:first-cluster} and~\ref{eq:second-cluster} we have that with probability at least $1-\nmo$:

	\begin{align}
		\min_{i\in[n]}d^*_i > c_4\log{n}
		\label{eq:fourth-concen}
	\end{align}
\end{proof}

\begin{lemma}
	\label{lemma:concen-high-prob}
	A graph $G$ generated by a Binary Asymmetric SBM with two communities sized $\rho n$ and $(1-\rho)n$ for some constant $\rho$ and with $p = \frac{a\log{n}}{n}$ and $q = \frac{b\log{n}}{n}$, there exists some tuples of constants $c_1, c_2, c_3, c_4$ such that $G$ is $(c_1, c_2, c_3, c_4)$-concentrated with probability at least $1 - \nmo$.
\end{lemma}

\begin{proof}
	In this analysis, we inherit some analyses from Lemma $3$ and Theorem $1$ of~\cite{hajek2016extensions}. We use different bounds on the Second and the Fourth conditions, which are stronger than similar bounds in~\cite{hajek2016extensions}.

	\textbf{The first condition} can be derived directly from~\cite{hajek2016achieving}'s Theorem 5. We adapt (\cite{hajek2016achieving}, Theorem 5) as Lemma~\ref{lemma:adj-mat-concen} for the convenience of our analyses. Since each edge of $G$ is generated with probability at least $b\log{n}/n$ and at most $a\log{n}/n$, by Lemma~\ref{lemma:adj-mat-concen}, with probability at least $1-\nmo$, there exists some constant $c_1$ such that:

	\begin{align}
		\Vert A - \E[A]\Vert_2 \leq c_1\sqrt{n\frac{\log{n}}{n}} = c_1\sqrt{\log{n}}
		\label{eq:first-concen}
	\end{align}

	Applying the union bound on the statement above and the results of Lemma~\ref{lemma:second-cond} (for \textbf{Second condition}), Lemma~\ref{lemma:third-cond} (\textbf{Third condition}), and Lemma~\ref{lemma:fourth-cond} (\textbf{Fourth condition}), there exists some tuples of constants $c_i$ such that G is $(c_1, c_2, c_3, c_4)$-concentrated with probability at least $1-\nmo$.

\end{proof}

\begin{lemma}
	\label{lemma:concen-stable}
	If $G$ is $(c_1, c_2, c_3, c_4)$-concentrated then for every graph $G': d(G, G') < \frac{c}{\epsilon}\log{n}$, i.e., $G'$ can be constructed by flipping at most $\frac{c}{\epsilon}\log{n}$ connections of $G$, $G'$ is $(c_1', c_2', c_3', c_4')$-concentrated, where $c_1' = c_1 + \sqrt{2c/\epsilon}, c_2' = c_2-c/\epsilon, c_3' = c_3+\sqrt{2c(1-\rho)/\epsilon\rho}, c_4' = c_4-c/\epsilon$.
\end{lemma}

\begin{proof}
	Given the graph $G$ is $(c_1, c_2, c_3, c_4)$-concentrated, it means that the graph $G$ satisfies all four conditions of the \emph{concentration} notation. We prove that all graph $G'$ that can be constructed by flipping up to $\frac{c\log{n}}{\epsilon}$ connections of $G$ will also satisfy all four conditions of \emph{concentration}, albeit with slightly different tuples of constants.

	\textbf{The first condition:} We follow~\cite{pmlr-v162-mohamed22a} to prove that $G'$ will satisfy the first condition. Let $\bar{A}$ be the expected adjacency matrix of graphs generated by the SBM. For both $G$ and $G'$ generated by the same SBM, we have $\E[A] = \E[A'] = \bar{A}$. Also, because $G'$ can be formed by flipping up to $ \frac{c\log{n}}{\epsilon}$ connections of $G$, then $\Vert A - A'\Vert_F \leq \sqrt{2c\log{n}/\epsilon}$. Now consider the $\ell_2$-norm of the difference between $A'$ and $\E[A']$:

	\begin{align}
		\Vert A' - \E[A'] \Vert_2 & = \Vert A' - \bar{A} \Vert_2                        \\
		                          & = \Vert A' - A + A - \bar{A} \Vert_2                \\
		                          & \leq \Vert A - \bar{A} \Vert + \Vert A' - A \Vert_2 \\
		                          & \leq c_1\sqrt{\log{n}} + \Vert A' - A \Vert_F       \\
		                          & \leq c_1\sqrt{\log{n}} + \sqrt{2c\log{n}/\epsilon}  \\
		                          & =(c_1 + \sqrt{2c/\epsilon})\sqrt{\log{n}}           \\
		                          & =c_1'\sqrt{\log{n}}.
	\end{align}

	\textbf{The second condition: } We revisit the second condition: $\xcheck^\T D^*\xcheck + \J(\xcheck) > c_2\log{n}$ holds for the graph $G$, where $\J(\xcheck) =\left(\lambda^* - \frac{p+q}{2}\right)\xcheck^\T J\xcheck $. By the definitions of the elements of the second conditions, $\xcheck, \lambda^*, p, q$ are all constants that are determined by the SBM and are not dependent on any instance of graph $G$ and $J$ is a constant matrix. Only the matrix $D^*$ is associated with and dependent on each instance of the graph $G$. Let $D^*, {D^*}'$ be the matrices that are associated with $G$ and $G'$, we have to prove that: $\xcheck^\T {D^*}'\xcheck + \J(\xcheck) > c_2'\log{n}$ for the graph $G'$.

	We have that:

	\begin{align}
		\xcheck^\T {D^*}'\xcheck + \J(\xcheck) & = \left(\xcheck^\T {D^*}'\xcheck - \xcheck^\T D^*\xcheck\right) + \left(\xcheck^\T D^*\xcheck +  \J(\xcheck)\right) \\
		                                       & > \xcheck^\T ({D^*}' - D^*)\xcheck  + c_2\log{n}.
		\label{eq:second-stability}
	\end{align}

	Now we analyze $\xcheck^\T ({D^*}' - D^*)\xcheck$. Let $\Delta^* = {D^*}' - D^*$, we have that $\Delta^*$ is also a diagonal matrix which $|\Delta^*_{ii}| = |{d^*}'_i - d^*_i| = |\sum_j(\Aij-\Aij')\sigma^*_i\sigma^*_j| \leq \frac{c\log{n}}{\epsilon}$, since there are at most $\frac{c\log{n}}{\epsilon}$ different entries between $\Aij$ and $\Aij'$ for any fixed $i$. Also, it is similar to check that $|\sum_i \Delta^*_{ii}| \leq 2c\log{n}/\epsilon$. We set $\mathbf{\Delta} =\{\Delta: |\sum_i \Delta_{ii}| \leq 2c\log{n}/\epsilon, |\Delta_{ii}| \leq c\log{n}/\epsilon\}$.

	Also, we have:

	\begin{align}
		\xcheck^\T ({D^*}' - D^*)\xcheck & \geq -|\xcheck^\T ({D^*}' - D^*)\xcheck|                                                                        \\
		                                 & \geq -\left|\max_{{D^*}'': d(G, G'')\leq c\log{n}/\epsilon}\xcheck^\T ({D^*}'' - D^*)\xcheck\right|             \\
		                                 & \geq -\left|\max_{\Delta\in \mathbf{\Delta}}\xcheck^\T \Delta \xcheck\right|                                    \\
		                                 & \overset{(a)}{\geq} -\left|\max_{\Delta\in \mathbf{\Delta}}\lambda_{max}(\Delta)\Vert \xcheck \Vert_2^2 \right| \\
		                                 & \overset{(b)}{\geq} -\left|\max_{\Delta\in \mathbf{\Delta}}\lambda_{max}(\Delta)\right|                         \\
		                                 & \overset{(c)}{\geq} -\frac{c\log{n}}{\epsilon},
	\end{align}

	where
    \begin{itemize}
        \item  $(a)$ is because of  $\xcheck^\T \Delta \xcheck \leq \lambda_{max}(\Delta)\Vert\xcheck\Vert_2^2$ for any symmetric $\Delta$;
        \item 	$(b)$ is because of $\Vert \xcheck \Vert = 1$ by its definition;
        \item 	and $(c)$ is because of $\Delta$ is a diagonal matrix so $\max_{\Delta}\lambda_{max}(\Delta) = \max_\Delta\max_i\Delta_{ii} \leq c\log{n}/\epsilon$. Adding this lower bound to Equation~\ref{eq:second-stability}, we have $\xcheck^\T {D^*}'\xcheck + \J(\xcheck) > (c_2 -c/\epsilon)\log{n} = c_2'\log{n}$.
    \end{itemize}

	\textbf{The third condition: } We will prove that $\Vert ({D^*}' - \E[{D^*}')\xcheck]\Vert_2 \leq c_3'\sqrt{\log{n}}$.

	We first analyze $\E[{D^*}']$. By the definition of ${D^*}'$, $\E[{D^*}'] = diag(\E[{d^*}'])$. For every $i$, we have $\E[{d^*}'_i] = \E[\sum_{j}\Aij\sigma^*_i\sigma^*_j - \lambda^*(2K-n)\sigma^*_i]$. By the linearity of expectation, we have:

	\begin{align}
		\E[{d^*}'_i] & = \sum_j\E[\Aij'] \sigma^*_i\sigma^*_j - \lambda^*(2K-n)\sigma^*_i \\
		             & = \sum_j\E[\Aij] \sigma^*_i\sigma^*_j - \lambda^*(2K-n)\sigma^*_i  \\
		             & = \E[d^*_i],
	\end{align}
	since we assume that all graphs $G$ and $G'$ are generated by the same SBM so each entry of their adjacency matrices must have the same expected value, and all other elements of the equations above are constants that only depend on the SBM's parameters. It follows that $\E[{D^*}'] = \E[D^*]$.

	Applying the property to $({D^*}' - \E[{D^*}')$:
	\begin{align}
		\Vert ({D^*}' - \E[{D^*}')\xcheck]\Vert_2 & = \Vert ({D^*}' - D^* + D^* - \E[{D^*})\xcheck \Vert_2                           \\
		                                          & \leq \Vert ({D^*}' - D^*)\xcheck \Vert_2 + \Vert (D^* - \E[{D^*})\xcheck \Vert_2 \\
		                                          & \leq \Vert ({D^*}' - D^*)\xcheck \Vert_2 + c_3\sqrt{\log{n}}
		\label{eq:mid-point-third-condition}
	\end{align}

	For the quantity $\Vert ({D^*}' - D^*)\xcheck \Vert_2$, using a similar analysis as in Lemma~\ref{lemma:third-cond}, we have;

	\begin{align}
		\Vert ({D^*}' - D^*)\xcheck \Vert_2
		 & =\sqrt{\sum_i{\xcheck_i^2\left(\sum_j(\Aij-\Aij')\sigma^*_j)\right)^2}}                                                          \\
		 & \leq\sqrt{\sum_i\left(\sum_j(\Aij-\Aij')\sigma^*_j \right)^2}\max\left\{ \sqrt{\frac{n-K}{nK}}, \sqrt{\frac{K}{n(n-K)}} \right\} \\
		 & \leq\Vert A - A'\Vert_F\max\left\{ \sqrt{\frac{1-\rho}{\rho}}, \sqrt{\frac{\rho}{1-\rho}} \right\}                               \\
		 & \leq \sqrt{\frac{2c\log{n}}{\epsilon}}\sqrt{\frac{1-\rho}{\rho}}                                                                 \\
		 & = \sqrt{\frac{2c(1-\rho)}{\epsilon\rho}}\sqrt{log{n}}.
	\end{align}

	Substituting it to Equation~\ref{eq:mid-point-third-condition}, we have:

	\begin{align}
		\Vert ({D^*}' - \E[{D^*}')\xcheck]\Vert_2 & \leq \sqrt{\frac{2c(1-\rho)}{\epsilon\rho}}\sqrt{log{n}} + c_3\sqrt{\log{n}}   \\
		                                          & \leq \left( \sqrt{\frac{2c(1-\rho)}{\epsilon\rho}} + c_3 \right)\sqrt{\log{n}} \\
		                                          & =c_3'\sqrt{\log{n}}
	\end{align}

	\textbf{The fourth condition: } For any graph $G'$, we prove that ${d^*}'_i \geq c_4'\log{n}$ for every $i\in[n]$. By the definition of ${d^*}'$, we have $\forall i$:

	\begin{align}
		{d^*}'_i & = \sum_j\Aij'\sigma^*_i\sigma^*_j - \lambda^*(2K-n)\sigma^*_i                                                                   \\
		         & = \sum_j\Aij'\sigma^*_i\sigma^*_j - \sum_j\Aij\sigma^*_i\sigma^*_j + \sum_j\Aij\sigma^*_i\sigma^*_j - \lambda^*(2K-n)\sigma^*_i \\
		         & =\sum_j\Aij'\sigma^*_i\sigma^*_j - \sum_j\Aij\sigma^*_i\sigma^*_j + d^*_i                                                       \\
		         & \geq \sum_j(\Aij'-\Aij)\sigma^*_i\sigma^*_j + c_4\log{n}                                                                        \\
		         & \overset{(a)}{\geq} -\frac{c\log{n}}{\epsilon} +c_4\log{n}                                                                       \\
		         & = \left(-\frac{c}{\epsilon}+c_4\right) \log{n}                                                                                   \\
		         & =c_4'\log{n},
	\end{align}

	where $(a)$ is because for each fixed $i$, $\Aij$s and $\Aij'$s differ by at most $c\log{n}/\epsilon$ entries across all $j$.

\end{proof}

\begin{lemma} (Lemma 3 of~\cite{hajek2016extensions})
	\label{lemma:binary-cert}
	Suppose there exist $D^* = diag\{d^*_i\}$ and $\lambda^*\in\mathbb{R}$ such that $S^* = D^* -A +\lambda^*J$ satisfies $S^*\curlyeqsucc 0, \lambda_2(S^*) > 0$ and $S^*\sigma^* = 0$.
	Then $Y^*$ is the unique solution of the program $SDP(G)$.
\end{lemma}

\begin{lemma}
	\label{lemma:sdp-optimal}
	If $G$ is $\C$-concentrated then $SDP(G) = Y^*$, i.e, the optimal solution of $SDP(G)$ is the ground truth community matrix $Y^* = \sigma^*\sigma^{*T}$.
\end{lemma}

\begin{proof}
	With $D^*$ and $\lambda^*$ defined as above, we show that when a graph $G$ is $\C$-concentrated, $S^* = D^* - A + \lambda^*J$ satisfies Lemma~\ref{lemma:binary-cert}'s requirements and the Lemma follows.

	We recall that

	\begin{itemize}
		\item $\tau = \frac{a-b}{\log{a}-\log{b}}$
		\item $\lambda^* = \tau\log{n}/n$
		\item $d^*\in\mathbb{R}^n: d^*_i = \sum_{j=1}^n\Aij\sigma^*_i\sigma^*_j - \lambda^*(2K - n)\sigma^*_i$
		\item $D^* = diag\{d^*\}$
	\end{itemize}

	For any $i$, we have $\sigma^*_i\sigma^*_i = 1$. Therefore, $d^*_i\sigma^*_i = \sum_{j=1}^n\Aij\sigma^*_i\sigma^*_j\sigma^*_i - \lambda^*(2K - n)\sigma^*_i\sigma^*_i = \sum_{j=1}^n\Aij\sigma^*_j - \lambda^*(2K - n)$. We have $D^*\sigma^* = A\sigma^* - \lambda^*(2K - n)\mathbf{1}$. It follows that $S^*\sigma^* = D^*\sigma^* - A\sigma^* +\lambda^*J\sigma^* = A\sigma^* - \lambda^*(2K - n)\mathbf{1} - A\sigma^* -\lambda^*J\sigma^* = 0$, which satisfies the condition that $S^*\sigma^* = 0$. Because of this, proving $\inf_{x\perp \sigma^*, \Vert x \Vert = 1} x^TS^*x > 0$ is sufficient to satisfy all remaining conditions, since all feasible $x$ plus $\sigma^*$ will include a basis for the whole space, which means $\forall y: y^TS^*y \geq 0$ ($S^* \curlyeqsucc 0$) and the solution set of $S^*y = 0$ has only $1$ dimension ($\lambda_2(S^*) > 0$). \\

	From now, we will show that when a graph $G$ is $\C$-concentrated,

	\begin{align}
		\inf_{x\perp \sigma^*, \Vert x \Vert = 1} x^TS^*x > 0
	\end{align}

	with probability $1$.


	Note that $\E[A] = \frac{p-q}{2}Y^* + \frac{p+q}{2}J - pI$. For any $x: x\perp \sigma^*, \Vert x \Vert = 1$:

	\begin{align}
		x^TS^*x & = x^TD^*x - x^TAx + x^T\lambda^*Jx                                                                              \\
		        & = x^TD^*x - x^T(A-\E[A])x - x^T\E[A]x+ x^T\lambda^*Jx                                                           \\
		        & = x^TD^*x - x^T(A-\E[A])x - \frac{p-q}{2}x^TY^*x - \frac{p+q}{2}x^TJx + px^TIx - \lambda^*x^TJx                 \\
		        & = x^TD^*x - x^T(A-\E[A])x - \frac{p-q}{2}x^T\sigma^*\sigma^{*T}x - \frac{p+q}{2}x^TJx + px^TIx - \lambda^*x^TJx \\
		        & \overset{(a)}{=} x^TD^*x - x^T(A-\E[A])x + (\lambda^* - \frac{p+q}{2})x^TJx +p\Vert x \Vert_2                   \\
		        & = x^TD^*x - x^T(A-\E[A])x + (\lambda^* - \frac{p+q}{2})x^TJx +p                                                 \\
		        & \overset{(b)}{\geq} p - \lambda_1(A-\E[A])\Vert x \Vert_2^2 + x^TD^*x + (\lambda^* - \frac{p+q}{2})x^TJx        \\
		        & = p - \lambda_1(A-\E[A])+ x^TD^*x + (\lambda^* - \frac{p+q}{2})x^TJx                                            \\
		        & \overset{(c)}{\geq} p - \Vert A-\E[A]\Vert_2 + x^TD^*x + (\lambda^* - \frac{p+q}{2})x^TJx                       \\
		        & \overset{(d)}{\geq} p - c_1\sqrt{\log{n}} + x^TD^*x + (\lambda^* - \frac{p+q}{2})x^TJx, \label{eq:semi-def}
	\end{align}

	where:

    \begin{itemize}
        \item (a) is because $x^T\sigma^* = 0$,
        \item (b) is because $x^TBx < \lambda_1(B)\Vert x \Vert_2^2$ for any symmetric matrix $B$,
        \item (c) is because $\lambda(B) < \Vert B \Vert_2$ for any matrix $B$,
        \item  and (d) is because of the \textbf{First condition} of $\C$-concentration.
    \end{itemize}

	Let $t(x) = x^TD^*x + (\lambda^* - \frac{p+q}{2})x^TJx$. By the definition of $\xcheck$ (Definition~\ref{def:binary}), we define $E = span(\xcheck, \sigma^*)$. Any $y: y\perp \sigma^*, \Vert y \Vert_2 = 1$ can be represented as $y = \beta\xcheck + \sqrt{1-\beta^2}x$ for $x\in\{x: x \perp E, \Vert x \Vert_2 = 1\}$ and $\beta\in[0, 1]$. We have:
	\begin{align}
		\inf_{x\perp \sigma^*, \Vert x \Vert_2 = 1} t(x) & = \inf_{x\perp E, \Vert x \Vert_2 = 1, \beta\in[0,1]}t(\beta\xcheck + \sqrt{1-\beta^2}x)                                                                                                      \\
		                                                 & \overset{(a)}{=}\inf_{x\perp E, \Vert x \Vert_2 = 1, \beta\in[0,1]}\left(\beta^2(\xcheck^TD^*\xcheck + \J(\xcheck)) + 2\beta\sqrt{1-\beta^2}x^TD^*\xcheck + (1-\beta^2)x^TD^*x\right)         \\
		                                                 & \geq \inf_{\beta\in{0,1}}(\beta^2(\xcheck^TD^*\xcheck + \J(\xcheck)) + \inf_{x\perp E, \Vert x \Vert_2 = 1, \beta\in[0,1]}(2\beta\sqrt{1-\beta^2}x^TD^*\xcheck + (1-\beta^2)x^TD^*x)          \\
		                                                 & \overset{(b)}{\geq}\inf_{\beta\in{0,1}}(\beta^2c_2\log{n} + \inf_{x\perp E, \Vert x \Vert_2 = 1, \beta\in[0,1]}(2\beta\sqrt{1-\beta^2}x^TD^*\xcheck + (1-\beta^2)x^TD^*x)                     \\
		                                                 & \overset{(c)}{\geq}\inf_{\beta\in{0,1}}(\beta^2c_2\log{n} - 2\beta\sqrt{1-\beta^2}\Vert (D^*-\E[D^*])\xcheck) \Vert  + \inf_{x\perp E, \Vert x \Vert_2 = 1, \beta\in[0,1]}(1-\beta^2)x^TD^*x) \\
		                                                 & \overset{(d)}{\geq}\inf_{\beta\in{0,1}}(\beta^2c_2\log{n} - 2\beta\sqrt{1-\beta^2}c_3\sqrt{\log{n}}  + \inf_{x\perp E, \Vert x \Vert_2 = 1, \beta\in[0,1]}(1-\beta^2)x^TD^*x)                 \\
		                                                 & \overset{(e)}{\geq}\inf_{\beta\in{0,1}}(\beta^2c_2\log{n} - 2\beta\sqrt{1-\beta^2}c_3\sqrt{\log{n}}  + (1-\beta^2)\min_i d^*_i)                                                               \\
		                                                 & \overset{(f)}{\geq}\inf_{\beta\in{0,1}}(\beta^2c_2\log{n} - 2\beta\sqrt{1-\beta^2}c_3\sqrt{\log{n}}  + (1-\beta^2)c_4\log{n})                                                                 \\
		                                                 & \geq\inf_{\beta\in{0,1}}(\beta^2c_2\log{n} + (1-\beta^2)c_4\log{n}) - c_3\sqrt{\log{n}}                                                                                                       \\
		                                                 & \geq\frac{1}{2}\min\{c_2, c_4\}\log{n} - c_3\sqrt{\log{n}}.
	\end{align}

	where:

	\begin{itemize}
		\item $(a)$ is because $Jx = 0$,
		\item $(b)$ is because $\xcheck^TD^*\xcheck + \J(\xcheck) \geq c_2\log{n}$ as the \textbf{second condition},
		\item $(c)$ is because $\inf_{x\perp E, \Vert x \Vert_2 = 1}x^TD^*\xcheck = \inf_{x\perp E, \Vert x \Vert_2 = 1} x^T(D^*-\E[D^*])\xcheck \geq -\Vert(D^* - \E[D^*]) \Vert$,
		\item $(d)$ is because $\Vert(D^* - \E[D^*]) \Vert < c_3\sqrt{\log{n}}$ as the \textbf{third condition},
		\item $(e)$ is because $D^*$ is a diagonal matrix constructed from $d_i^*$,
		\item $(f)$ is because $min_i d_i^* \geq c_4\log{n}$ as the \textbf{fourth condition}.
	\end{itemize}

	Substituting the lower bound of $t(x)$ to Equation~\ref{eq:semi-def}, we have:

	\begin{align}
		x^TS^*x & \geq\frac{1}{2}\min\{c_2, c_4\}\log{n} - (c_1+c_3)\sqrt{\log{n}} + p \\
		        & > 0,
	\end{align}

	where $n$ is large enough.

\end{proof}

\begin{lemma}
	\label{lemma:stability1}
	A graph $G$ generated by a Binary Asymmetric SBM with two communities sized $\rho n$ and $(1-\rho)n$ for some constant $\rho$ and with $p = \frac{a\log{n}}{n}$ and $q = \frac{b\log{n}}{n}$ is $\frac{c}{\epsilon}\log{n}$-stable under $SDP(G)$ with probability at least $1-\nmo$
\end{lemma}

\begin{proof}
	By Lemma~\ref{lemma:concen-high-prob}, $G$ is $\C$-concentrated with probability at least $1-\nmo$. By Lemma~\ref{lemma:concen-stable}, all graphs $G'$ (which include $G$ itself) with distance up to $\frac{c}{\epsilon}\log{n}$ from $G$ are also $\C'$-concentrated. By Lemma~\ref{lemma:sdp-optimal}, since all graphs $G'$ are $C'$-concentrated, $SDP(G')$ always output the optimal and unique solution $\sigma^*$ with probability $1$, or $SDP(G) = SDP(G') = \sigma^*$ for all $G'$. It follows that $G$ is $\frac{c}{\epsilon}\log{n}$-stable under $SDP$ with probability at least $1-\nmo$.
\end{proof}

\begin{theorem}
   \label{theorem:basbm}
	Given a graph $G$ generated by a Binary Asymmetric SBM with two communities sized $\rho n$ and $(1-\rho)n$ for some constant $\rho$, and with $p = \frac{a\log{n}}{n}$ and $q = \frac{b\log{n}}{n}$, $\tilde{h}(c/\epsilon) > 1$, $\sqrt{a} - \sqrt{b(1+\log{\frac{a}{b}})} > \sqrt{c\log{\frac{a}{b}}/\epsilon}$, $M_{SDP}$ with $\delta = n^{-\Omega(1)}$ exactly recovers the ground-truth community $\sigma^*$, i.e., $\Pr[M^{SDP}(G)\neq \sigma^*] = \nmo$
\end{theorem}

\begin{proof}
	Lemma~\ref{lemma:stability1} states the stability property of $G$ under $SDP$, i.e., $G$ is $\frac{c}{\epsilon}\log{n}$-stable under $SDP$. It also implies that $SDP(G)=\sigma^*$ under the effect of $\mathcal{C}$-concentration. By applying Lemma~\ref{lemma:stability-mechanism}, $\Pr[M^{SDP}(G) \neq SDP(G)] = \nmo$ or $\Pr[M^{SDP}(G) \neq \sigma^*] = \nmo$.

	The condition for the Theorem is that $\C'$ is a valid constant combination, i.e., $c_2' > 0$ and $c_4' > 0$, or $c_2 - c/\epsilon > 0$ and $c_4 - c/\epsilon > 0$ with pre-determined $c, \epsilon$.

	Lemma~\ref{lemma:fourth-cond} requires that $\tilde{h}(c_4) >1$, which means $\tilde{h}(c/\epsilon) > 1$



	Lemma~\ref{lemma:second-cond} requires that $c_2 < \tau - b$, which means $c/\epsilon < \tau -b$.
    This is equivalent to $a - b(1+\log{\frac{a}{b}}) > c\log{\frac{a}{b}}/\epsilon$, which can be simplified to $\sqrt{a} - \sqrt{b(1+\log{\frac{a}{b}})} > \sqrt{c\log{\frac{a}{b}}/\epsilon}$.
\end{proof}

%
%
%
%
%


\section{Censored Binary SBM (CBSBM)}

In this section we analyze the Stability mechanism on the recoverability of the Censored Binary SBM. In this model, the generated graphs are bedge-weighted. The vertices are consists of nodes from two clusters with possibly unequal sizes. Edges between these nodes are generated by an Erdos-Renyi model $G(n,p)$, regardless of the endpoint's communities. With a fixed constant $\xi\in[0,0.5]$, each edge $(i,j)$ has a label $L_{ij}\in \{+1, -1\}$ drawn from the following distribution:

\begin{definition} Definition of CBSBM.
  \label{def:cbsbm}

\begin{align}
  P_{L_{ij}|\sigma^*_i,\sigma^*_j} &= (1-\xi)\bm{1}_{L_{ij} = \sigma^*_i\sigma^*_j} + \xi \bm{1}_{L_{ij} = -\sigma^*_i\sigma^*_j}
\end{align}
\end{definition}

\textbf{Privacy model.}
The adjacency matrix $A(G)$ is defined as $A_{ij} = 0$ if there is no edge between $i$ and $j$. $A_{ij} = L_{ij}$ if there is an edge generated between $i$ and $j$.
In this section, we define the neighborhood between two graph $G\sim G'$ if $A(G) and A(G')$ differ by exact one element.
This privacy model can protect the existence (and the non-existence) of an arbitrary edge $(i,j)$, where any two neighboring adjacency matrices differ at element $ij$: $A(G)_{ij} = 0$ (not an edge) and $A(G')_{ij} \neq 0$ (an edge).
It can also protect the label of an arbitrary edge $(i,j)$ whenever $(i,j)$ exists in the input graphs, that any two neighboring adjacency matrix differ as follows: $A(G)_{ij} = -1$ and $A(G')_{ij} = +1$.

In the non-private setting, communities in the CBSBM can be recovered exactly by solving the following SDP Relaxation:

\begin{definition}
  SDP Relaxation of the Censored Binary SBM:
  \begin{align}
    \hat{Y}_{SDP} &= \argmax \langle A, Y \rangle \\
    \text{s.t.   } Y &\curlyeqsucc 0 \\
    Y_{ii} &= 1, i \in [n].
  \end{align}
\end{definition}

\textbf{Assumptions of parameters.}
We assume that $p = a\log{n}/n$ for some fixed constant $a$ (in the random edge generation model $G(n,p)$). For the random label model: $h(\xi, a) = a(\sqrt{1-\xi} -\sqrt{\xi})^2 > 1$, or $h(\xi, a) = 1 + \Omega(1)$. We may drop parameter $a$ when the context is clear (CBSBM with a fixed $a$).

In this section, we denote the $SDP(G)$ as a function taking input graph $G$ and outputting the optimal solution of the SDP Relaxation constructed by its adjacency matrix $A(G)$.

\begin{definition} In this section, we define the following quantities:
  \begin{itemize}
  \item $d^*_i = \sum_{j = 1}^n \Aij\sigma^*_i\sigma^*_j$ for every $i \in [n]$
  \item $D^* = diag\{d^*\}$
  \end{itemize}
\end{definition}

\begin{definition}
  \label{def:cbsbm-concen}
  Definition of $C$-concentration.
  Assume a graph $G$ is generated by a CBSBM with two communities with ground truth vector $\sigma^*$, edge probability $p = \frac{a\log{n}}{n}$, and edge labels are generated by the above process with some constant $\xi \in [0, 0.5]$.

  $G$ is called $\mathcal{C}$-concentrated if there exists a tuple $(c_1, c_2)$ such that $G$ satisfies two conditions
  \begin{itemize}
  \item $\Vert A - \E[A]\Vert_2 \leq c_1 \sqrt{\log{n}}$
  \item $d^*_i \geq c_2 \log{n}$  for every $i\in [n]$
  \end{itemize}
\end{definition}

\begin{lemma}
  \label{lemma:censored-first-high-prob}
  (Theorem 9 of\cite{hajek2016extensions})
  Given a graph $G$ generated by a CBSBM as in Definition~\ref{def:cbsbm}, there exists a constant $c_1 > 0$ such that with probability at least $1-\nmo$, we have:
  \begin{align}
    \Vert A - \E[A]\Vert_2 &\leq c_1 \sqrt{\log{n}}
  \end{align}
\end{lemma}

\begin{lemma}
  \label{lemma:binom-tail}
  (Lemma 1 of~\cite{hajek2016extensions})
  Let $X~\sim Binom(m, a\log{n}/n)$ for $m\in \mathbb{N}$ where $m = \rho n + o(n)$ for some $\rho > 0$ as $n\rightarrow \infty$. Let $k_n \in [m]$ be such that $k_n = \tau\rho\log{n} + o(\log{n})$ for some $0 \leq \tau \leq a$, then:
  \begin{align}
    \Pr[X \leq k_n] &= n^{\rho(a - \tau\log\frac{ea}{\tau} + o(1))}.
  \end{align}
\end{lemma}

\begin{lemma}
  \label{lemma:censored-second-high-prob}
Given a graph $G$ generated by a CBSBM as in Definition~\ref{def:cbsbm}, there exists a constant $0 < c_2 < a$ such that with probability at least $1-\nmo$, we have:
  \begin{align}
    \min_{i\in [n]} d^*_i &\geq c_2\log{n}
  \end{align}
\end{lemma}

\begin{proof}
  The proof here is inspired by~\cite{hajek2016extensions}, with the main differences are the bound in the original proof is in order of $\frac{\log{n}}{\log{\log{n}}}$, where we need a bound in order of $\log{n}$ for the remaining proof of Stability to work.
  By definition, $d^*_i = \sum_{j = 1}^n\Aij\sigma^*_i\sigma^*_j$. In the generation process, it is equivalent to $d^*_i = \sum_{j=1}^{n-1}X_j$, where $X_j \overset{i.i.d.}{\sim} p(1-\xi)\beta_{+1} + p\xi\beta_{-1} + (1-p)\beta_0$ where $\beta_x$ is the Dirac delta function at $x$. Hence, we will prove that with probability at least $1-\nmo$, there exist some constant $c_2 > 0$ that for every $i$:

  \begin{align}
    \Pr[\sum_{j=1}^{n}X_j < c_2\log{n}] &\leq \nmo 
  \end{align}

  We first analyze the case when $\xi = 0$. Then $\sum_{j=1}^{n}X_j \sim Binom(n, a\log{n}/n)$. It follows that:

  \begin{align}
    \Pr[\sum_{j=1}^{n}X_j < c_2\log{n}] &\leq n^{-\rho(a -\tau\log{\frac{ea}{\tau}} + o(1))} \text{ by Lemma~\ref{lemma:binom-tail}}\\
    &= n^{-a + c_2\log{\frac{ea}{c_2}} - o(1)} \text{ by substituting $\rho = 1, \tau = c_2$}\\
    &\leq n^{-a - o(1)} \text{ since $c_2 < a$}\\
    &\leq n^{-h(\xi)} \text{ since $h(\xi) = a$ when $\xi = 0$}\\
    &\leq n^{-1 - \Omega(1)}.
  \end{align}

  Taking the union bound over all nodes $i$, we have $d^*_i \geq c_2\log{n}$ with probability at least $1-\nmo$.

  Now we analyze the case when $\xi > 0$. By the Chernoff bound:
  \begin{align}
    \label{eq:cheroff}
    \Pr[\sum_{j=1}^{n}X_j < c_2\log{n}] &\leq  \exp(-n\ell(\frac{c_2\log{n}}{n})),
  \end{align}

  where the function $\ell(x)$ is defined as $\ell(x) = \sup_{\lambda \geq 0}-\lambda x - \log{\E[e^{-\lambda X}]}$ with $X \sim p(1-\xi)\beta_{+1} + p\xi\beta_{-1} + (1-p)\beta_0$, or $\E[e^{-\lambda X}] = 1 + p(e^{-\lambda}(1-\xi) + e^\lambda\xi - 1)$. Since $\ell(x)$ is concave (in $\lambda$), the supremum at $\lambda^*$ is:

  \begin{align}
    -x + \frac{p(e^{-\lambda^*}(1-\xi) - e^{\lambda^*}\xi}{1 + p(e^{-\lambda^*}(1-\xi) + e^{\lambda^*}\xi - 1)} = 0.
  \end{align}

  Substituting $x = \frac{c_2\log{n}}{n} \approx 0$, we solve $\lambda^*$ as follows:

  \begin{align}
    e^{-\lambda^*}(1-\xi) - e^{\lambda^*}\xi  &= o(1) \\
    \implies e^{\lambda^*}(e^{-2\lambda^*}(1-\xi)-xi) &= o(1)\\
    \implies e^{-2\lambda^*} &= \frac{\xi}{1-\xi} + o(1) \\
    \implies -2\lambda^* &= \log{\frac{\xi}{1-\xi}} + o(1) \\
    \implies \lambda^* &= \frac{1}{2}\log{\frac{1-\xi}{\xi}} + o(1).
  \end{align}
  
  Substituting $x = \frac{c_2\log{n}}{n}$ and $\lambda^* = \frac{1}{2}\log{\frac{1-\xi}{\xi}} + o(1)$, we have:

  \begin{align}
    \ell(\frac{c_2\log{n}}{n}) &= -\lambda^*\frac{c_2\log{n}}{n} - \log(1 + p(e^{-\lambda^*}(1-\xi) + e^{\lambda^*}\xi - 1))\\
                               &= -\frac{1}{2} \log\frac{1-\xi}{\xi}\times \frac{c_2\log{n}}{n} - \log(1-p(\sqrt{1-\xi} - \sqrt{\xi})^2) + o(\frac{\log{n}}{n})\\
    &\lessapprox - \log(1-p(\sqrt{1-\xi} - \sqrt{\xi})^2) + o(\frac{\log{n}}{n}), \text{ since }\frac{c_2\log{n}}{n} \gtrapprox 0 \\
                               &=- p(\sqrt{1-\xi} - \sqrt{\xi})^2 + o(\frac{\log{n}}{n}), \text{ since }\log(1-x) = -x \text{ at }x \approx 0\\
    &= \frac{a\log{n}}{n}(\sqrt{1-\xi} - \sqrt{\xi})^2 + o(\frac{\log{n}}{n}).
  \end{align}

  Applying the above result to Equation~\ref{eq:cheroff}, we have:
  \begin{align}
    \Pr[\sum_{j=1}^{n}X_j < c_2\log{n}] &\leq  \exp(-n\ell(\frac{c_2\log{n}}{n}))\\
&\leq\exp(-n \times \frac{a\log{n}}{n}(\sqrt{1-\xi} - \sqrt{\xi})^2 - n\times o(\frac{\log{n}}{n})) \\
&\leq\exp(-a\log{n}(\sqrt{1-\xi} - \sqrt{\xi})^2 - o(\log{n})) \\
&\leq n^{-h(\xi) - \Omega(1)} \\
&\leq n^{-1 - \Omega(1)}.
  \end{align}

  Taking the union bound over all nodes $i$, we have $d^*_i \geq c_2\log{n}$ with probability at least $1-\nmo$ and the Lemma follows.
\end{proof}

\begin{lemma}
  \label{lemma:cbsbm-concen-high-prob}
A graph $G$ generated by a CBSBM as in Definition~\ref{def:cbsbm}, there exists some tuple of constants $(c_1, c_2)$ such that with probability at least $1-\nmo$, $G$ is $(c_1, c_2)$-concentrated ($\mathcal{C}$-concentrated).
\end{lemma}

\begin{proof}
Refer to the definition of $\mathcal{C}$-concentrated for CBSBM (Definition~\ref{def:cbsbm-concen}), the \textbf{First condition} follows from Lemma~\ref{lemma:censored-first-high-prob} and the \textbf{Second condition} follows from Lemma~\ref{lemma:censored-second-high-prob}. Applying union bound on both conditions, we have that with probability at least $1-\nmo$, $G$ is $\mathcal{C}$-concentrated.
\end{proof}

\begin{lemma}
  \label{lemma:cbsbm-distance-logn}
Assume a fixed CBSBM models, if a graph $G$ generated by a CBSBM is $(c_1, c_2)$-concentrated then every graph $G'$ within distance $\frac{c\log{n}}{\epsilon}$ of $G$, i.e. $d(G,G') < \frac{c\log{n}}{\epsilon}$, is $(c'_1, c'_2)$-concentrated where $c_1' = c_1 + \sqrt{8c/\epsilon}$, $c_2' = c_2 - c/\epsilon$.
\end{lemma}

\begin{proof}
  
  \textbf{The first condition: }
  It is clear that $A(G')$ can be obtained by changing up to $\frac{c\log{n}}{\epsilon}$ cells of $A(G)$, with the largest impact (to the concentration properties) by changing some cell $ij$ from $A(Gx)_{ij} = 1$ to $A(G')_{ij} = -1$ or vice versa. It follows that $\Vert A(G) - A(G') \Vert^2_F \leq \frac{8c\log{n}}{\epsilon}$. Also, since we assume $G$ and $G'$ are under the same CBSBM, $\E[A(G')] = \E[A(G)] = \bar{A}$.

  Now we analyze $\Vert A(G') - \E[A(G')]\Vert_2$:

  \begin{align}
    \Vert A(G') - \E[A(G')]\Vert_2 &= \Vert A(G') - \bar{A} \Vert_2 \\
    &= \Vert A' - A + A - \bar{A} \Vert_2\\
    &\leq \Vert A' - A \Vert_2 + \Vert A - \bar{A} \Vert_2\\
    &\leq \Vert A' - A \Vert_2 + c_1\sqrt{\log{n}}\\
    &\leq \sqrt{\frac{8c\log{n}}{\epsilon}} + c_1\sqrt{\log{n}}\\
                                   &=(\sqrt{8c/\epsilon } + c_1)\sqrt{\log{n}}\\
    &=c'_1\sqrt{\log{n}},
  \end{align}
  which means that $G'$ satisfies the \textbf{First condition} with some constant $c'_1$.

  \textbf{The first condition: }
  By the definition of $d'^*_i$, we have:

  \begin{align}
    d'^*_i &= \sum_{j = 1}^{n}\Aij'\sigma^*_i\sigma^*_j \\
           &= \sum_{j = 1}^{n}(\Aij' - \Aij)\sigma^*_i\sigma^*_j + \sum_{j = 1}^{n}\Aij\sigma^*_i\sigma^*_j\\
           &\geq   \sum_{j = 1}^{n}(\Aij' - \Aij)\sigma^*_i\sigma^*_j + c_2\log{n}\\
           &\geq -\frac{2c\log{n}}{\epsilon} + c_2\log{n}\\
    &= (c_2 - \frac{2c}{\epsilon})\log{n}\\
    &=c_2'\log{n},
  \end{align}

  which means that $G'$ satisfies the \textbf{Second condition} with some constant $c'_2$, and the Lemma follows.
\end{proof}

\begin{lemma}
  \label{lemma:cbsbm-cert}
  (Lemma 9 of\cite{hajek2016extensions})
 Suppose there exists $D^*=diag{d^*_i}$ such that $S^* = D^* -A$ satisfies $S^* \curlyeqsucc 0, \lambda_2(S^*) > 0$ and $S^*\sigma^* = 0$. Then $Y^*$ is the unique solution of the program $SDP(G)$.
\end{lemma}

\begin{lemma}
  \label{lemma:cbsbm-optimal}
  If $G$ is $\C$-concentrated then $SDP(G) = Y^*$, i.e., the optimal solution of $SDP(G)$ is the ground truth community matrix $Y^* = sigma^*\sigma^{*T}$.
\end{lemma}

\begin{proof}
  Constructing $d^*_i = \sum_{j=1}^n\Aij\sigma^*_i\sigma^*_j$, we will prove that $S^*$ satisfies the condition of Lemma~\ref{lemma:cbsbm-cert} and due to Lemma~\ref{lemma:cbsbm-cert}, the Lemma follows that $Y^*$ is the optimal solution of $SDP(G)$.

  First, we prove that $S^*\sigma^* = 0$. By definition of $S^* = D^* - A$, $S^*\sigma^* = D^*\sigma^* - A\sigma^* = A^*\sigma^* - A^*\sigma^* = 0$

  Second, proving that $\inf_{x\perp \sigma^*, \Vert x\Vert = 1}x^TS^*x > 0$ satisfies other conditions since all feasible $x$ plus $\sigma^*$ will include a basis for the whole space, which means $\forall y: y^TS^*y \geq 0$, or $S^*\curlyeqsucc 0$ and the solution set of $S^*y = 0$ has only 1 dimension $\lambda_2(S^*) > 0$.

  Therefore, the next step is to prove that when $G$ is $\C$-concentrated, $\inf_{x\perp \sigma^*, \Vert x\Vert = 1}x^TS^*x > 0$ with probability $1$ and the proof follows. For any $x\perp \sigma^*, \Vert x \Vert = 1$, we have:

  \begin{align}
    x^TS^*x &= x^TD^*x - x^TAx \\
            &= x^TD^*x - x^T(A - \E[A])x -  x^T\E[A]x\\
            &\overset{(a)}{\geq} \min_{i\in[n]}d^*_i - \Vert A - \E[A] \Vert_2 - x^T\E[A]x\\
            &\overset{(b)}{\geq} c_2\log{n} - c_1\sqrt{\log{n}} - x^T\E[A]x\\
            &\overset{(c)}{=}c_2\log{n} - c_1\sqrt{\log{n}} + (1-2\xi)px^TY^*x\\
            &=c_2\log{n} - c_1\sqrt{\log{n}} + (1-2\xi)p\\
            &=c_2\log{n} - c_1\sqrt{\log{n}} + (1-2\xi)a\log{n}/n\\
            &\geq 0 \text{ where $n$ is large enough,}
  \end{align}
  
  where:
  \begin{itemize}
  \item $(a)$ is because of $\Vert x \Vert = 1$,
  \item $(b)$ is because of the $\C$-concentration's conditions,
  \item $(c)$ is because of $\E[A] = (1-2\xi)Y^*$.
  \end{itemize}
    
\end{proof}

\begin{lemma}
	\label{lemma:cbsbm-stability-final}
A graph $G$ generated by a CBSBM as in Definition~\ref{def:cbsbm} is $\frac{c}{\epsilon}\log{n}$-stable under $SDP(G)$ with probability at least $1-\nmo$.
\end{lemma}

\begin{proof}
	By Lemma~\ref{lemma:cbsbm-concen-high-prob}, $G$ is $\C$-concentrated with probability at least $1-\nmo$. By Lemma~\ref{lemma:cbsbm-distance-logn}, all graphs $G'$ (which include $G$ itself) with distance up to $\frac{c}{\epsilon}\log{n}$ from $G$ are also $\C'$-concentrated. By Lemma~\ref{lemma:cbsbm-optimal}, since all graphs $G'$ are $C'$-concentrated, $SDP(G')$ always output the optimal and unique solution $\sigma^*$ with probability $1$, or $SDP(G) = SDP(G') = Y^*$ for all $G'$. It follows that $G$ is $\frac{c}{\epsilon}\log{n}$-stable under $SDP$ with probability at least $1-\nmo$.
\end{proof}

\begin{theorem}
  \label{threorem:cbsbm}
Given graph $G$ generated by a CBSBM as in Definition~\ref{def:cbsbm} 
, and $c/\epsilon < a$ and $h(\xi, a) > 1, c/\epsilon < a$, $M_{SDP}$ with $\delta = n^{-\Omega(1)}$ exactly recovers the ground-truth community $Y^*$, i.e., $\Pr[M^{SDP}(G)\neq Y^*] = \nmo$
\end{theorem}

\begin{proof}
	Lemma~\ref{lemma:cbsbm-stability-final} states the stability property of $G$ under $SDP$, i.e., $G$ is $\frac{c}{\epsilon}\log{n}$-stable under $SDP$. It also implies that $SDP(G)=Y^*$ under the effect of $\mathcal{C}$-concentration. By applying Lemma~\ref{lemma:stability-mechanism}, $\Pr[M^{SDP}(G) \neq SDP(G)] = \nmo$ or $\Pr[M^{SDP}(G) \neq Y^*] = \nmo$.

	The condition for the Theorem is that $\C'$ is a valid constant combination, i.e., $c_2' < a$ and hence, $h(\xi, a) > 1$ and $ c/\epsilon < a$.
\end{proof}


\section{Polynomial Algorithms}

In Algorithm~\ref{alg:stability-mechanism}, there are two computation tasks: computing $d_{SDP}(G)$ and $SDP(G)$. Solving the SDP Relaxation can be done in polynomial-time. Therefore, the remaining question is how long it takes to compute $d_{SDG}(G)$. One way is to start with all neighbors $G'$ of distance $k=1$ of $G$ and test if $SDP(G) \overset{?}{=} SDP(G')$. If all $G'$ satisfy the test, we can conclude that $G$ is at least $1$-stable, otherwise $G$ is unstable. If $G$ is unstable, we increase $k$ by one and repeat the procedure until we find the first $G''$ that has $SDP(G'') \neq SDP(G)$ and $d_{SDP}(G) = k$. We may apply the trick of~\cite{pmlr-v162-mohamed22a} that stops when $k = c\log{n}/\epsilon$. In this case, because there are $n^{O(\log{n})}$ neighbors of distance up to $c\log{n}/\epsilon$ from $G$, we have to invoke the solver for the SDP Relaxation $n^{O(\log{n})}$ times. Since $d_{SDP}(G) = O(\log{n})$  w.h.p., Algorithm~\ref{alg:stability-mechanism} takes $n^{O(\log{n})}$ w.h.p..

The main idea is if we can estimate $d(G)$ faster, we can design a faster algorithm. We note that if the input graph is $\C$-concentrated, then $d(G) \geq c\log{n}/\epsilon$. Therefore, we can test if the input graph is $\C$-concentrated. If the test comes out as positive, we can set $\widehat{d}(G) = d(G) \geq c\log{n}/\epsilon$ and use $\widehat{d}$ instead of $d$. Else, we compute $\widehat{d}(G) = \min(d(G), c\log{n}/\epsilon)$. It is clear that w.h.p., the test is positive. The main challenge is that, testing $\C$-concentration requires knowledge of the SBMs, i.e., $p$, $q$ and \textbf{most importantly}, $\sigma^*$ (or $\xi^*_k$ in $r > 2$ communities)--the quantities we are trying to output. 

In several settings of applications, the edge probabilities $p$ and $q$ (and therefore $a, b$ respectively) may be known, which makes the problem easier. Generally, it is safe to assume that we do not know $a$ and $b$. To construct the conditions of $\C$-concentration, we need a reliable way to estimate them from the input graph.

Suppose that we have access to oracles that can provide us these quantities. Let $\Oracle_{\sigma^*}$ be the one that can provide us the true value of $\sigma^*$ (or $\xi^*_k$ respectively). Let $\Oracle_{a,b}^\alpha$ be the one that can provide us the parameters $\hat{a}, \hat{b}$ accurately up to a factor of $1\pm\alpha$ from the true values of $a, b$ for a small constant $\alpha < 0.001$. 

We present Algorithm~\ref{alg:oracle-mechanism} with the unrealistic assumption of the oracles. We then prove that Algorithm~\ref{alg:oracle-mechanism} retains the privacy and utility of Algorithm~\ref{alg:stability-mechanism}, which means it is private and achieve exact recovery. Because checking the $\C$-concentration can be done in polynomial time, and w.h.p., we do not have to invoke high computational cost $d(G)$, the Algorithm takes polynomial time. After we confirm that $\M_{\Stbl\Oracle}$ has all the properties we need, we will replace the oracles by realistic components that we calculate from the input graph $G$. We then present Algorithm~\ref{alg:poly-stability-mechanism} that w.h.p. is the same with Algorithm~\ref{alg:oracle-mechanism}.

In order to do that, we employ an estimator derived from the results of~\cite{hajek2016extensions} and formalize it in Algorithm~\ref{alg:estimate}.

\begin{algorithm}
  \caption{$\M^{f}_{\Stbl\Oracle}(G, \C)$: Fast Stability Mechanism with \Oracle}
  \label{alg:oracle-mechanism}
  \begin{algorithmic}[1]
    \STATE $\hat{Y} \leftarrow f(G)$
    \STATE $\sigma^* \leftarrow \Oracle_{\sigma^*}$
    \STATE $(\hat{a}, \hat{b}) \leftarrow \Oracle^\alpha_{a,b}$
    \STATE $\hat{\C} \leftarrow \text{ adjust }\C \text{ on }\alpha \text{ to satisfy Proposition~\ref{prop:oracle-concen}}$
    \STATE Construct $\hat{\C}$-concentration using $\hat{C}, \sigma^*, \hat{a}, \hat{b}$
    \IF {$G$ is $\hat{\C}$-concentrated}
    \STATE $\hat{d}(G) \leftarrow c\log{n}/\epsilon$
    \ELSE
    \STATE $\hat{d}(G) \leftarrow \min(c\log{n}/\epsilon, d(G))$
    \ENDIF
    \STATE $\tilde{d}(G) \leftarrow \hat{d}(G) + \Lap(1/\epsilon)$
    \IF{$\tilde{d}_f(G)  > \frac{\log{1/\delta}}{\epsilon}$}
    \STATE Output $\hat{Y}$
    \ELSE
    \STATE Output $\perp$ 
    \ENDIF
  \end{algorithmic}
\end{algorithm}


 In each SBM setting, the proposition can be easily verified by checking all conditions of $\C$-concentration. $\Oracle_{\sigma^*}$~guarantees us the true value of $\sigma^*$, so the differences between $\C$ and $\hat{\C}$ only come from the factor $\alpha$ of $\Oracle_{a,b}^\alpha$. We use a tighter tuple of constants $\hat{\C}$ (compared to $\C$) to balance the fact that $\hat{a}$ and $\hat{b}$ may be off by some factor of $1\pm \alpha$. This task can be done by adjust each condition by scaling the respective $c_k$ to a factor of $1\pm 2\alpha$ in which direction that makes the condition tighter.

%

 \begin{lemma}
   \label{lemma:oracle-privacy}
Algorithm~\ref{alg:oracle-mechanism} is $(\epsilon,\delta)$-differentially private.
 \end{lemma}

 \begin{proof}
   Suppose that $\Delta_{\hat{d}} = 1$ (the global sensitivity of $\hat{d}$). Using the same arguments in the proof of Theorem~\ref{theorem:privacy}, substituting $d$ by $\hat{d}$, it follows that the algorithm is $(\epsilon, \delta)$-differentially private. It remains that we need to prove $\Delta_{\hat{d}} = 1$.

   Suppose we have a pair of neighbors $G\sim G'$. By definition $\Delta_{\hat{d}} = \max_{G\sim G'}|\hat{d}(G) - \hat{d}(G')|$. If both of them pass the test in line 6, $\hat{d}(G) = \hat{d}(G') = c\log{n}/\epsilon$ which means $|\hat{d}(G) - \hat{d}(G')| = 0 < 1$. If both of them fail the test in line 6, $\hat{d}(G) = d(G)$ and $\hat{d}(G') = d(G)$, which means $|\hat{d}(G) - \hat{d}(G')| = |d(G) - d(G')| \leq 1$.

   In the last case, assume that $G$ passes the test in line 6 while $G'$ fails. It means that $\hat{d}(G) = c\log{n}/\epsilon$ and $\hat{d}(G') = d(G')$. Suppose $d(G') \leq c\log{n}/\epsilon - 2$, because $\Delta_d = 1$ and $G\sim G'$, $d(G) \leq d(G') + 1 \leq c\log{n}/\epsilon - 1$. But it contradicts with the fact that $G$ is $\hat{\C}$-concentrated which implies that that $G$ is $c\log{n}/\epsilon$-stable or $d(G) \geq c\log{n}/\epsilon$. It shows that $d(G') > c\log{n}/\epsilon - 2$ or $\hat{d}(G') \geq c\log{n}/\epsilon$ or $\hat{d}(G) - \hat{d}(G') \leq 1$ and the Lemma follows.
 \end{proof}

 \begin{lemma}
  \label{lemma:oracle-exact-recovery} 
  If $\M^{f}_{\Stbl}$ (Algorithm~\ref{alg:stability-mechanism}) achieves exact recovery (under some specific conditions of the SBMs under the view of Lemma~\ref{lemma:prelim-stability-mechanism}), $\M^f_{\Stbl\Oracle}$ (Algorithm~\ref{alg:oracle-mechanism}) also achieves exact recovery under the same conditions.
 \end{lemma}
 
 \begin{proof}
   Since $\hat{\C}$-concentration is a valid concentration, it follows that $\Pr[G\text{ is } \hat{\C}\text{-concentrated}] \geq 1-\nmo$. Also, since $\hat{C}$-concentration implies $\C$-concentration by Proposition~\ref{prop:oracle-concen}, and all graphs satisfies $\C$-concentration are $c\log{n}/\epsilon$-stable in view of the assumed SBM. It follows that all graphs that satisfies $\hat{C}$-concentration are at least $c\log{n}/\epsilon$-stable in view of the assumed SBM. Applying Lemma~\ref{lemma:prelim-stability-mechanism}, $\M^f_{\Stbl\Oracle}$ (Algorithm~\ref{alg:oracle-mechanism}) achieves exact recovery in the same conditions.
 \end{proof}

 \begin{lemma}
   \label{lemma:oracle-poly}
If $\M^{f}_{\Stbl}$ (Algorithm~\ref{alg:stability-mechanism}) achieves exact recovery (under some specific conditions of the SBMs under the view of Lemma~\ref{lemma:prelim-stability-mechanism}), Algorithm~\ref{alg:oracle-mechanism} takes $O(poly(n))$ times w.h.p..
 \end{lemma}

 \begin{proof}
   Using the same arguments as in Lemma~\ref{lemma:oracle-exact-recovery}, the input graph $G$ satisfies $c\log{n}/\epsilon$-stable in view of the assumed SBM. It means that w.h.p., $\hat{d}(G)$ is set to $c\log{n}/\epsilon$ in line 6 instead of going through the computation of $d(G)$ in line 9. Since checking the concentration conditions takes $O(poly(n))$ times, and solving the SDP Relaxation at line 1 takes $O(poly(n))$ times, it follows that w.h.p., Algorithm~\ref{alg:oracle-mechanism} takes $O(poly(n))$.
 \end{proof}

\begin{lemma}
 \label{lemma:param-estimate} 
 (Proof in Appendix B of~\cite{hajek2016extensions})
 Given $G$ generated by a BASBM with $\rho \leq 0.5$, Algorithm~\ref{alg:estimate} outputs $(\ahat, \bhat ) = (a, b) + o(1)$ w.h.p..
\end{lemma}

\begin{lemma}
  \label{lemma:poly}
  If $\M^{f}_{\Stbl}$ (Algorithm~\ref{alg:stability-mechanism}) achieves exact recovery (under some specific conditions of the SBMs under the view of Lemma~\ref{lemma:prelim-stability-mechanism}), Algorithm~\ref{alg:poly-stability-mechanism} ($\M^{f}_{\Stbl \textsc{Fast}}$) is the same with Algorithm~\ref{alg:oracle-mechanism} ($\M^{f}_{\Stbl\Oracle}$) w.h.p..
\end{lemma}

\begin{proof}
Algorithm~\ref{alg:poly-stability-mechanism} is identical with Algorithm~\ref{alg:oracle-mechanism}, except in two steps that Algorithm~\ref{alg:oracle-mechanism} invokes the oracles.

For the first oracle in line 2, we replace $\Oracle_{\sigma^*}$ by $\hat{Y}$, which is our estimation of $\sigma^*$ by the solving the SDP Relaxation. It is clear that under our assumption of exact recovery, $\hat{Y} =\sigma^*(\sigma^*)^T$ w.h.p..  Therefore using $\hat{Y}$ is as good as asking \Oracle w.h.p.

For the second oracle in line 3, we replace $\Oracle^\alpha_{a,b}$ by $\ahat, \bhat$ estimated by Algorithm~\ref{alg:estimate}, derived from~\cite{hajek2016extensions}. Due to Lemma~\ref{lemma:param-estimate}, $(\ahat, \bhat) = (a,b) + o(1)$ w.h.p.. It means that when $n$ is large enough, $\ahat, \bhat$ is at least as good as $\Oracle^\alpha_{a,b}$ and the Lemma follows.

\end{proof}


\begin{algorithm}
  \caption{$ParamEstimate(G)$: Algorithm to estimate $a, b$ for BASBM}
  \label{alg:estimate}
  \begin{algorithmic}[1]
    \STATE Let $A$ be the adjacency matrix of the input graph $G$
    \STATE $d_{i} \leftarrow \sum_{j}A_{ij}$
    \STATE $w_i \leftarrow \frac{d_i}{\log{n}}$
    \STATE $\what \leftarrow \frac{1}{n}\sum_{i}w_i$
    \STATE $\hat{\rho} \leftarrow \frac{1}{n}\sum\bm{1}_{\{w_i \leq \what\}}$
    \STATE $\what_+ \leftarrow \frac{1}{n}\sum w_i\bm{1}_{\{w_i \geq \what\}}$
    \STATE $\what_- \leftarrow \frac{1}{n}\sum w_i\bm{1}_{\{w_i \leq \what\}}$
    \STATE $\ahat \leftarrow \frac{(1-\rhohat)\what_+ - \rhohat\what_-}{1-2\rhohat}$ 
    \STATE $\bhat \leftarrow \frac{(1-\rhohat)\what_- - \rhohat\what_+}{1-2\rhohat}$ 
    \STATE \textbf{Return } $\bm{\ahat, \bhat}$
  \end{algorithmic}
\end{algorithm}


\section{General Structure SBM (GSSBM)}
\label{sec:gssbm}

In the General Structure SBM (GSSBM), there are multiple possibly unequal communities (i.e., $r > 0$ clusters) and some outliers.
The cluster $k^{th}, k \in [r]$, or $C_{k}$ has size $K_k = \rho_k\times n$ as $n\rightarrow \infty$. For any $i > j$, assume that $\rho_i \geq \rho_j > 0$.
The are $n-\sum_{k\in[r]}K_k$ outlier vertices do not belong to any cluster. We use $k=0$ (e.g., in $C_0, K_0$) to refer to the outliers.
Similar to other SBM, we consider the dense regime, where edges are generated with probability $\Omega(\log{n}/n)$.
The ground truth community matrix $Z^* = \sum_{k\in[r]}\xi^*_k(\xi_k)^T$ where $\xi_k$ is the indicator vector of community $C_k$.

\begin{definition}
  \label{def:gssbm}
A graph $G$ with $r$ communities indicated by vectors $\xi_i, i\in[r]$. Edges whose endpoints from a same cluster are generated with probability $p = \frac{a\log{n}}{n}$ and other edges are generated with probability $q = \frac{b\log{n}}{n}$ for some constants $a > b >0$.
\end{definition}

In the non-private setting, the community matrix $Z$ can be obtained by solving the following SDP Relaxation. We use $SDP(G)$ to denote the function taking input $G$ and outputting the optimal solution of the SDP Relaxation.

\begin{definition}
  \label{def:gssbm-sdp}
  The SDP Relaxation of GSSBM.
  \begin{align}
    \hat{Z}_{SDP} &= \argmax \langle A, Z \rangle \\
    \text{ s.t. } Z &\succcurlyeq 0 \\
    Z_{ii} &\leq 1 \text{ for } i \in [n]\\
    Z_{ij} &\geq 0 \text{ for } i,j \in [n]\\
    \langle I, Z\rangle &= \sum_{k =1}^r K_k \\
    \langle J, Z\rangle &= \sum_{k=1}^r K_k^2
  \end{align}
  
\end{definition}

\begin{definition}
  We define some quantities used in our analyses:
  \begin{itemize}
  \item $A(G)$ is the adjacency matrix of graph $G$. We drop the parameter when the context is clear.
  \item $e(i, C_k)$ is the number of edges between a node $i$ and nodes from cluster $C_k$
  \item $k(i)$ is $i$'s cluster
  \item $s_i = e(i,C_{k(i)})$
  \item $\tilde{\tau} = b + 2c_2$
  \item $\lambda^* = \tilde{\tau}\log{n}/n$
  \item $d^*_i = 
    \begin{cases}
      s_i - \Vert A - \E[A] \Vert_2 - \lambda^*K_k \text{ for }i \in C_k, k \in [r]\\
      0 \text{ for }i\in C_0
    \end{cases}$ 
  \item $D^* = diag\{d^*\}$
  \item $B^* \in \mathcal{S}^n, B^*_{ij} =
    \begin{cases}
      \lambda^* + \frac{e(C_{k(i)}, C_{k(j)})}{K_{(k(i))}K_{(k(j))}} - \frac{e(i, C_{k(j)})}{K_{k(j)}} - \frac{e(j, C_{k(i)})}{K_{k(i)}} \text{ for } k(i) \neq k(j), k(i), k(j) \in [r]\\
      \lambda^*  - \frac{e(i, C_{k(j)})}{K_{k(j)}} \text{ for } k(i) = 0, k(j) \in [r]\\
      \lambda^*  - \frac{e(j, C_{k(i)})}{K_{k(i)}} \text{ for } k(j) = 0, k(i) \in [r]\\
      0 \text { for } k(i) = k(j)
    \end{cases}
   $ 
 \item $E_r = span\{\xi_1, \xi_2, \cdots, \xi_r\}$
   \item $\E[A]$ is a constant matrix that denotes the expected values of $A$ over the randomness of the SBM process. $\E[A] = (p-q)Z^* -pI^{(in)} - qI^{(out)} + qx^TJx$
   \item $\rho_{k, k\in[r]} = K_k/n$
   \item $\rho_{min} = min_{k\in[r]}\rho_k$
   \item $I(x,y) = x - y\log{\frac{ex}{y}}$
  \end{itemize}
\end{definition}

\textbf{Assumptions of parameters}

\begin{align}
  &I(a, b+2c_2) > 1/\rho_{min} \text{ or } I(a,b + 2c_2) = 1/\rho_{min}+ \Omega(1)\\
  &I(b, b+c_2-c_3/\rho_{min}) > 1/\rho_{min} \text{ or } I(b,b + c_2 - c_3/\rho_{min}) = 1/\rho_{min} + \Omega(1)\\
  &I(b, b+2c_2-c_5/\rho_{min}) > 1/\rho_{min} \text{ or } I(b,b + 2c_2 - c_5/\rho_{min}) = 1/\rho_{min}+ \Omega(1)
\end{align}

\begin{definition}
  \label{def:gssbm-concen}
  Definition of $\C$-concentration for GSSBM.
  Given a graph $G$ generated by a $GSSBM$ defined is Definition~\ref{def:gssbm}. $G$ is called $\C$-concentrated if there exists a tuple of constants $(c_1, c_2, c_3, c_4, c_5)$ such that $G$ satisfies these conditions:
  \begin{itemize}
  \item $\Vert A(G) - \E[A(G)] \Vert \leq c_1\sqrt{\log{n}}$
  \item $\min_{i\in [n]}s_i \geq (b + 2c_2)\rho_{k(i)}\log{n}$
  \item $\max_{i\in[n], k: k \neq k(i)}e(i, C_{k}) \leq (b + c_2)K_{k}\log{n}/n - c_3\log{n}$
  \item $\min_{i,j: k(i)k(j)[k(i)-k(j)] \neq 0}e(C_{k(i)}, C_{k(j)})\geq K_{k(i)}K_{k(j)}q - 2\sqrt{K_{k(i)}K_{k(j)}}\sqrt{{\log{n}}} - c_4\log{n}$
  \item $\max_{i\in C_0}e(i, C_{k: k \neq 0}) \geq \tilde{\tau}K_r\frac{\log{n}}{n} - c_5\log{n}$
  \end{itemize}
\end{definition}

\begin{lemma}
  \label{lemma:gssbm-second-concen-high-prob}
  Given a graph $G$ generated by a $GSSBM$ defined is Definition~\ref{def:gssbm}. There exists some constant $c_2$ such that:
  \begin{align}
\min_{i\in [n]}s_i \geq (b + 2c_2)\rho_{k(i)}\log{n}
  \end{align}
 with probability at least $1-\nmo$.
\end{lemma}

\begin{proof}
  Since $s_i$ is the number of edges between $i$ and vertices from the same cluster with $i$, we have $s_i \sim Binom(K_{k(i)}, \frac{a\log{n}}{n})$, where $K_{k(i)} = \rho_{k(i)}n$. Applying Lemma~\ref{lemma:binom-tail}, with $k_n = (b+2c_2)\log{n}$, we have:

 \begin{align}
   \Pr[s_i \leq (b + 2c_2)\rho_{k(i)}\log{n}] &=  n^{-\rho_{k(i)}(a - (b+2c_2)\log{\frac{ea}{b + 2c_2}} + o(1))}\\
                                              &\leq n^{-\rho_{k(i)}I(a, b+2c_2)}\\
                                              &\leq n^{-1 -\Omega(1)},
 \end{align}
 where the first inequality is because of the definition of $I$ and the last inequality is because of the assumption of parameters. Taking union bound on all vertices $i$ and the Lemma follows.
  
\end{proof}

\begin{lemma}
  \label{lemma:gssbm-third-concen-high-prob}
  Given a graph $G$ generated by a $GSSBM$ defined is Definition~\ref{def:gssbm}. There exists some constant $c_2, c_3$ such that:
  \begin{align}
\max_{i\in[n], k: k \neq k(i)}e(i, C_{k}) \leq (b + c_2)K_{k} - c_3\log{n},
  \end{align}
 with probability at least $1-\nmo$.
\end{lemma}

\begin{proof}
 By the definition of $e(i, C_{k: k\neq k(i)})$, it is the number of edges of a vertex $i$ and vertices from a different cluster $k \neq k(i)$, or $e(i, C_{k: k\neq k(i)}) \sim Binom(K_k, \frac{b\log{n}}{n})$.  Applying Lemma~\ref{lemma:binom-tail}, with $k_n = (b+c_2 - c_3/\rho_k)\log{n}$, we have:

 \begin{align}
   \Pr[e(i, C_k) \geq k_n] &= \Pr[e(i, C_k) \geq (b+c_2 - c_3/\rho_k)\log{n}] \\
                           &\leq n^{-\rho_k(b - (b+c_2 -c_3/\rho_k)\log{\frac{eb}{b+c_2-c_3/\rho_k}} + o(1))}\\
                           &\overset{(a)}\leq n^{-\rho_kI(b, b+c_2-c_3/\rho_k)}\\
                           &\overset{(b)}\leq n^{-1-\Omega(1)},
 \end{align}

 where:

 \begin{itemize}
   \item $(a)$ is because of the definition of $I$,
   \item $(b)$ is because of the assumption of parameters.
 \end{itemize}

  Taking union bound on all vertices $i$ and the Lemma follows.
\end{proof}

\begin{lemma}
  \label{lemma:gssbm-fourth-concen-high-prob}
  Given a graph $G$ generated by a $GSSBM$ defined is Definition~\ref{def:gssbm}. There exists some constant $c_4$ such that:
  \begin{align}
\min_{i,j: k(i)k(j)[k(i)-k(j)] \neq 0}e(C_{k(i)}, C_{k(j)})\geq K_{k(i)}K_{k(j)}q - 2\sqrt{K_{k(i)}K_{k(j)}}\sqrt{{\log{n}}} - c_4\log{n},
  \end{align}
 with probability at least $1-\nmo$.
\end{lemma}

\begin{proof}
  Let $k = k(i), k' = k(j)$. The condition implies that $k$ and $k'$ are two cluster and not outliers.

  By the definition, $e(C_k, C_{k'}) = Binom(K_kK_{k'}, q)$. By applying standard Chernoff bound, we have:

  \begin{align}
	\Pr[e(C_k, C_k') \leq (1-\alpha)K_kK_{k'}\times q] \leq e^{-\alpha^2K_kK_{k'}q/2}.
    \label{eq:fifth}
  \end{align}

  Setting,
  \begin{align}
    \alpha = \frac{2\sqrt{K_kK_{k'}}\sqrt{\log{n}} + c_4\log{n}}{qK_kK_{k'}},
  \end{align}

  because $|c_4\log{n}| << \sqrt{K_kK_{k'}}\sqrt{\log{n}}$,

  \begin{align}
    \alpha &> \frac{\sqrt{K_kK_{k'}}\sqrt{\log{n}}}{qK_kK_{k'}}\\
    &= \frac{\log{n}}{q\sqrt{K_kK_{k'}}}.
  \end{align}

  Substituting $\alpha$ to Equation~\ref{eq:fifth}, we have:

  \begin{align}
 \Pr[e(C_k, C_{k'}) \leq qK_kK_{k'} - 2\sqrt{\log{n}}\sqrt{K_kK_{k'}} -c_4\log{n}]   &=  \Pr[e(C_k, C_{k'}) \leq (1-\alpha)K_kK_{k'}\times q]\\
                                                         &\leq e^{-(\frac{\sqrt{\log{n}}}{q\sqrt{K_kK_{k'}}})^2K_kK_{k'}q/2} \\
                                                         &= e^{-\frac{\log{n}}{2q}}\\
                                                                                     &=n^{-\frac{1}{2q}}\\
                                                                                      &< n^{-2 - \Omega(1)}, \text{ where } n \text{ is large enough},
  \end{align}

  Taking the union bound on all $k, k'$, the Lemma follows.
\end{proof}

\begin{lemma}
  \label{lemma:gssbm-fifth-concen-high-prob}
  Given a graph $G$ generated by a $GSSBM$ defined is Definition~\ref{def:gssbm}. There exists some constant $c_5$ such that:
  \begin{align}
  \max_{i\in C_0}e(i, C_{k(j)}) \geq \tilde{\tau}K_r\frac{\log{n}}{n} - c_5\log{n},
  \end{align}
 with probability at least $1-\nmo$.
\end{lemma}

\begin{proof}
 By the definition of $e(i, C_{k, k\neq 0})$, it is the number of edges of an outlier $i$ and vertices from a cluster $k \neq 0$, or $e(i, C_k) \sim Binom(K_k, \frac{b\log{n}}{n})$.  Applying Lemma~\ref{lemma:binom-tail}, with $k_n = (\tilde{\tau} - c_3/\rho_k)\log{n}$, we have:

 \begin{align}
   \Pr[e(i, C_k) \geq k_n] &= \Pr[e(i, C_k) \geq (\tilde{\tau} - c_5/\rho_k)\log{n}] \\
                           &\leq n^{-\rho_k(b - (\tilde{\tau} -c_5/\rho_k)\log{\frac{eb}{\tilde{\tau}-c_5/\rho_k}} + o(1))}\\
                           &\overset{(a)}\leq n^{-\rho_kI(b, b+2c_2-c_5/\rho_k)}\\
                           &\overset{(b)}\leq n^{-1-\Omega(1)},
 \end{align}

 where:

 \begin{itemize}
   \item $(a)$ is because of the definition of $I$,
   \item $(b)$ is because of the assumption of parameters.
   \end{itemize}
   
 Taking union bound on all vertices $i$ and the Lemma follows.
\end{proof}

\begin{lemma}
  \label{lemma:gssbm-concen-high-prob}
  Given a graph $G$ generated by a $GSSBM$ defined is Definition~\ref{def:gssbm}. There exists some tuples of constants $\C = (c_1, c_2, c_3, c_4, c_5)$ such that, $G$ is $\C$-concentrated,  with probability at least $1-\nmo$.
\end{lemma}

\begin{proof}
  The \textbf{First condition} can be proved using the same arguments and settings as in Lemma~\ref{lemma:concen-high-prob}, in which the edges of $G$ are generated with probabilities $\Omega(\log{n}/n)$. It means that the exists some constant $c_1$ such that $\Vert A - \E[A]\Vert_2 \leq c_1\sqrt{\log{n}}$ with probability at least $1-\nmo$. Using union bound on it and the \textbf{Second condition} (Lemma~\ref{lemma:gssbm-second-concen-high-prob}), \textbf{Third condition} (Lemma~\ref{lemma:gssbm-third-concen-high-prob}), \textbf{Fourth condition} (Lemma~\ref{lemma:gssbm-fourth-concen-high-prob}), and the \textbf{Fifth condition} (Lemma~\ref{lemma:gssbm-fifth-concen-high-prob}), there exists some tuples of constants $\C = (c_1, c_2, c_3, c_4, c_5)$ such that $G$ satisfies all the conditions with probability at least $1-\nmo$. It means that with probability $1-\nmo$, $G$ is $\C$-concentrated.
\end{proof}

\begin{lemma}
	\label{lemma:gssbm-concen-stable}
 Given a graph $G$ generated by a $GSSBM$ defined is Definition~\ref{def:gssbm}. 
 If $G$ is $\C$-concentrated then for every graph $G': d(G, G') < \frac{c}{\epsilon}\log{n}$, i.e., $G'$ can be constructed by flipping at most $\frac{c}{\epsilon}$ connections of $G$, $G'$ is $\C' = (c_1', c_2', c_3', c_4', c_5')$-concentrated, where

 \begin{align}
   c_1' &= c_1 + \sqrt{2c/\epsilon}\\
   c_2' &= c_2-\frac{c}{\epsilon\rho_{min}}\\
   c_3' &= c_3-\frac{c}{\epsilon}\\
   c_4' &= c_4 + c/\epsilon\\
   c_5' &= c_5 -\frac{c}{\epsilon}.
 \end{align}
\end{lemma}

\begin{proof}

 The \textbf{First condition} follows from the same arguments in Lemma~\ref{lemma:concen-stable}'s First condition. We will prove the remaining conditions.

 We denote $s'(), e'()$ as the same $s(), e()$ but with the graph $G'$ instead of $G$. We note that $G$ and $G'$ differ by at most $c\log{n}/\epsilon$ edges.

 The \textbf{Second condition}, we have:

 \begin{align}
   s'_i &\geq s_i - \frac{c\log{n}}{\epsilon}\\
    &\geq (b + 2c_2)\rho_{k(i)}\log{n} -\frac{c\log{n}}{\epsilon}\\
    &= (b+2c_2 - \frac{c}{\epsilon\rho_{k(i)}}) \rho_{k(i)}\log{n}\\
    &\geq (b+2c_2 - \frac{c}{\epsilon\rho_{min}}) \rho_{k(i)}\log{n}\\
   &= (b+2c'_2) \rho_{k(i)}\log{n}, 
 \end{align}

which implies $\min_{i\in [n]}s'_i \geq (b + 2c'_2)\rho_{k(i)}\log{n}$.

The \textbf{Third condition}, we have:

\begin{align}
  e'(i, C_{k}) &\leq e(i, C_{k}) + \frac{c\log{n}}{\epsilon}\\
               &\leq (b + c_2)K_{k}\log{n}/n - c_3\log{n}  + \frac{c\log{n}}{\epsilon}\\
               &\leq (b + c_2)K_{k}\log{n}/n - (c_3- \frac{c}{\epsilon})\log{n}  \\
               &=(b + c_2)K_{k}\log{n}/n - (c_3')\log{n},
\end{align}

which implies $\max_{i\in[n], k: k \neq k(i)}e'(i, C_{k}) \leq (b + c_2)K_{k}\log{n}/n - c'_3\log{n}$.

The \textbf{Fourth condition}, we have:

\begin{align}
e'(C_{k(i)}, C_{k(j)}) &\geq e(C_{k(i)}, C_{k(j)}) - \frac{c\log{n}}{\epsilon}\\
  &\geq K_{k(i)}K_{k(j)}q - 2\sqrt{K_{k(i)}K_{k(j)}}\sqrt{{\log{n}}} - c_4\log{n} - \frac{c\log{n}}{\epsilon}\\
  &\geq K_{k(i)}K_{k(j)}q - 2\sqrt{K_{k(i)}K_{k(j)}}\sqrt{{\log{n}}} - (c_4 + \frac{c}{\epsilon})\log{n} \\
  &\geq K_{k(i)}K_{k(j)}q - 2\sqrt{K_{k(i)}K_{k(j)}}\sqrt{{\log{n}}} - c_4'\log{n},
\end{align}

which implies $\min_{i,j: k(i)k(j)[k(i)-k(j)] \neq 0}e'(C_{k(i)}, C_{k(j)})\geq K_{k(i)}K_{k(j)}q - 2\sqrt{K_{k(i)}K_{k(j)}}\sqrt{{\log{n}}} - c_4\log{n}$.

The \textbf{Fifth condition}, we have:

\begin{align}
  e'(i, C_{k}) &\leq e(i, C_{k}) + \frac{c\log{n}}{\epsilon}\\
               &\leq (b + 2c_2)K_{k}\log{n}/n - c_5\log{n}  + \frac{c\log{n}}{\epsilon}\\
               &\leq (b + 2c_2)K_{k}\log{n}/n - (c_5- \frac{c}{\epsilon})\log{n}  \\
               &=(b + 2c_2)K_{k}\log{n}/n - (c_5')\log{n}\\
               &=\tilde{\tau}K_{k}\log{n}/n - (c_5')\log{n},
\end{align}
which implies $\max_{i\in C_0}e'(i, C_{k: k\neq 0}) \geq \tilde{\tau}K_r\frac{\log{n}}{n} - c'_5\log{n}$, and complete the proof of the Lemma.
\end{proof}

\begin{lemma}
  \label{lemma:gssbm-di}
  Given graph $G$ generated by the GSSBM in Definition~\ref{def:gssbm} that is $\C$-concentrated, $d^*_i > 0$ for every node $i$ belongs to a community, i.e., $k(i) \neq 0$.
\end{lemma}

\begin{proof}
Recall that by our choice, $d^*_i$ is define as:

\begin{align}
  d^*_i = 
    \begin{cases}
      s_i - \Vert A - \E[A] \Vert_2 - \lambda^*K_k \text{ for }i \in C_k, k \in [r]\\
      0 \text{ for }i\in C_0
    \end{cases}
\end{align}

where $\lambda^* = (b + 2c_2)\frac{\log{n}}{n}$. Because $G$ is $\C$-concentrated, $A-\E[A] \leq c_1\sqrt{\log{n}}$ because of the \textbf{First condition} and $\min_{i\in [n]}s_i \geq \lambda^*K_{k(i)} + c_2\log{n}$  because of the \textbf{Second condition}. We have for node $i: k(i) \neq 0$:

\begin{align}
  d^*_i &= s_i - \Vert A - \E[A] \Vert_2 - \lambda^*K_{k(i)}\\
        &\geq \lambda^*K_{k(i)} + c_2\log{n} - c_1\sqrt{\log{n}} -\lambda^*K_{k(i)}\\
        &=c_2\log{n} - c_1\sqrt{\log{n}} \\
  &> 0 \text{ with } n \text{ large enough,}
\end{align}

and the Lemma follows.
\end{proof}

\begin{lemma}
  \label{lemma:gssbm-bij}
  Given graph $G$ generated by the GSSBM in Definition~\ref{def:gssbm} that is $\C$-concentrated, $B^*_{ij} > 0$ for every vertices $i$ and $j$ belong to different communities, i.e., $k(i) \neq k(j)$.
\end{lemma}

\begin{proof}
  Recall that $B^* \in \mathcal{S}^n, B^*_{ij} =
    \begin{cases}
      \lambda^* + \frac{e(C_{k(i)}, C_{k(j)})}{K_{(k(i))}K_{(k(j))}} - \frac{e(i, C_{k(j)})}{K_{k(j)}} - \frac{e(j, C_{k(i)})}{K_{k(i)}} \text{ for } k(i) \neq k(j), k(i), k(j) \in [r]\\
      \lambda^*  - \frac{e(i, C_{k(j)})}{K_{k(j)}} \text{ for } k(i) = 0, k(j) \in [r]\\
      \lambda^*  - \frac{e(j, C_{k(i)})}{K_{k(i)}} \text{ for } k(j) = 0, k(i) \in [r]\\
      0 \text { for } k(i) = k(j)
    \end{cases}
   $, then we only care about the first three cases. 

   In the first case, since $G$ is $\C$-concentrated, we have $e(i, C_{k(j)}) \leq (b + c_2)K_{k(j)} - c_3\log{n}$ due to the \textbf{Third condition} and $e(C_{k(i)}, C_{k(j)})\geq K_{k(i)}K_{k(j)}q - 2\sqrt{K_{k(i)}K_{k(j)}}\sqrt{{\log{n}}} - c_4\log{n}$ due to the \textbf{Fourth condition}, we have:
l m  
\begin{align}
B^*_{ij} &= \lambda^* + \frac{e(C_{k(i)}, C_{k(j)})}{K_{(k(i))}K_{(k(j))}} - \frac{e(i, C_{k(j)})}{K_{k(j)}} - \frac{e(j, C_{k(i)})}{K_{k(i)}} \\
         &\geq \frac{\tilde{\tau}\log{n}}{n} - \frac{(b + c_2)K_{k_{min}} - c_3\log{n}}{K_{k(i)}} - \frac{(b + c_2)K_{k_{min}} - c_3\log{n}}{K_{k(i)}} \\
         &+ \frac{q\times K_{k(i)}K_{k(j)} - 2\sqrt{K_{k(i)}K_{k(j)}}\sqrt{\log{n}} - c_4\log{n}}{K_{k(i)}K_{k(j)}}\\
  &\geq(b + 2c_2 + c_3/\rho_{min} - b - c_2 + c_3/\rho_{min} - c_2 + b)\frac{\log{n}}{n} - \frac{2\sqrt{\log{n}}}{\rho_{min}n} - \frac{c_4\log{n}}{\rho_{min}n^2}\\
         &=\frac{2c_3}{\rho_{min}}\frac{\log{n}}{n} - \frac{2\sqrt{\log{n}}}{\rho_{min}n} - \frac{c_4\log{n}}{\rho_{min}^2n^2} \\
  &> 0 \text{  when } n \text{ is large enough}.
\end{align}
   
The second and third cases are similar. W.L.O.G., let $k(i) = 0, k(j) \neq 0$, we have:

\begin{align}
B^*_{ij} &= \lambda^*  - \frac{e(i, C_{k(j)})}{K_{k(j)}}\\
         &\geq \frac{(b + 2c_2)\log{n}}{n} - \frac{\tilde{\tau}K_r\log{n}}{\rho_{min}n^2} + \frac{c_5\log{n}}{\rho_{min}n}\\
         &= \frac{c_5\log{n}}{\rho_{min}n}\\
  &> 0,
\end{align}
where the first inequality is because of the \textbf{Fifth condition} of $\C$-concentration that $G$ satisfies as the assumption of the Lemma,
which completes the proof and the Lemma follows.
\end{proof}

\begin{lemma}
  \label{lemma:gssbm-cert}
  (Lemma~10 of~\cite{hajek2016extensions}) Suppose there exists $D^* = diag\{d^*_i\}$ with $d^*_i > 0$ for non-outlier vertices and $d^*_i = 0$ for outlier vertices, $B^* = \mathcal{S}^n$ with $B^* \geq 0$ and $B^*_{ij} > 0$ whenever $i$ and $j$ are in different clusters, $\eta^*\in\R$ and $\lambda^*\in\R$ such that $S^* \overset{def}{=} D^* - B^* -A + \eta^*I + \lambda^*J$ satisfies that $S^*\succcurlyeq 0$, $\lambda_{r+1}(S^*) > 0$ where $\lambda_r(S^*)$ is the $r^{th}$ smallest eigenvalue of $S^*$, and
  \begin{align}
    S^*\xi_k^* &= 0 \text{  for } k\in[r], \\
    B^*_{ij}Z^*_{ij} &=0 \text{  for } i,j\in[n].
  \end{align}

  Then $Z^*$ is the unique solution to the SDP in Definition~\ref{def:gssbm-sdp}
\end{lemma}

\begin{lemma}
  \label{lemma:gssbm-optimal}
  If a graph $G$ generated by the GSSBM in Definition~\ref{def:gssbm} is $\C$-concentrated, then $SDP(G) = Z^*$, i.e., the ground truth community matrix $Z^*$ is the optimal solution of the SDP Relaxation in Definition~\ref{def:gssbm-sdp}.
\end{lemma}

\begin{proof}
  We now prove that we can construct $S^*$ that satisfies the requirement of Lemma~\ref{lemma:gssbm-cert}. By that, we provide a certificate deterministically constructed from the $\C$-concentration property that guarantees that the optimal solution SDP Relaxation is the ground-truth community matrix $Z^*$. In other words, solving the SDP in Definition~\ref{def:gssbm-sdp} when the input graph $G$ is $\C$-concentrated gives us the ground-truth community matrix $Z^*$ with probability $1$.

  Lemma~\ref{lemma:gssbm-di} and Lemma~\ref{lemma:gssbm-bij} show us a way to construct $D^*$ and $B^*$ that satisfies Lemma~\ref{lemma:gssbm-cert}. The remaining part is to prove that $S^* \succcurlyeq 0$. It is equivalent to prove that $x^TS^*x \geq 0$ for $x\in \R^n$ and $x \perp E_r$, where $E_r = span\{\xi_1, cdots, \xi_r\}$.

  It is clear that:

  \begin{align}
    x^TB^*x &= \sum_{k\neq k'}\sum_{i\in C_k}\sum_{j\in C_{k'}} B^*_{ij}x_ix_j\\
            &= 0.
              \label{eq:B-star}
  \end{align}

  Now we analyze $x^T\E[A]x$. With $\E[A] = (p-q)Z^* - pI^{(in)} -qI^{(out)}- qJ$ where $I^{(in)}$ and $I^{(out)}$ are the identity matrices for in-cluster nodes and outliers, i.e., $I^{(in)}_{ii} = 1$ for every $k(i) \neq 0$ and $I^{(out)}_{ii} = 1$ for every $k(i) = 0$. We have:

  \begin{align}
    x^T\E[A]x &= (p-q)x^TZ^*x - px^TI^{(in)}x -qx^TI^{(out)}x + qx^TJx\\
    &= -p\sum_{i: k(i) \neq 0}x_i^2 -q\sum_{i: k(i) = 0}x_i^2 + qx^TJx,
      \label{eq:E[A]}
  \end{align}

  where the last equality is because $x^TZ^*x = 0$ for every $x \perp E_r$.

  By the definition of $S^* = D^* - B^* -A + \eta^*I + \lambda^*J$, choosing $\eta = \Vert A - \E[A]\Vert_2$, we have:

  \begin{align}
    x^TS^*x &= x^TD^*x - x^TB^*x - x^TAx + \eta^* x^TIx +\lambda^* x^TJx\\
    &\overset{(a)}{=} x^TD^*x - x^TAx + x^T(A - \E[A])x +\lambda^* x^TJx\\
    &\geq x^TD^*x - x^T\E[A]x + (b+2c_2)\frac{\log{n}}{n}x^TJx\\
    &\overset{(b)}{\geq} x^TD^*x + p\sum_{i: k(i) \neq 0}x_i^2  + q\sum_{i: k(i) = 0}x_i^2 - \frac{b\log{n}}{n}x^TJx + (b+2c_2)\frac{\log{n}}{n}x^TJx\\
    &\geq x^TD^*x + p\sum_{i: k(i) \neq 0}x_i^2  + q\sum_{i: k(i) = 0}x_i^2 + 2c_2\frac{\log{n}}{n}x^TJx\\
    &\geq x^TD^*x + p\sum_{i: k(i) \neq 0}x_i^2  + q\sum_{i: k(i) = 0}x_i^2 + 2c_2\frac{\log{n}}{n}(\sum_{i}x_i)^2\\
    &\geq x^TD^*x \\
    &\overset{(c)}{\geq} 0,
  \end{align}

  where:
  \begin{itemize}
    \item $(a)$ is because $x^TB^*X = 0$ for $x \perp E_r$  (Equation~\ref{eq:B-star}) and $\eta^* = \Vert A - \E[A]\Vert_2$,
    \item $(b)$ is because of Equation~\ref{eq:E[A]},
    \item $(c)$ is because of the result of Lemma~\ref{lemma:gssbm-di};
  \end{itemize}
  
  and the Lemma follows.
\end{proof}

\begin{lemma}
	\label{lemma:gssbm-stability-final}
A graph $G$ generated by a GSSBM as in Definition~\ref{def:gssbm} is $\frac{c}{\epsilon}\log{n}$-stable under $SDP(G)$ with probability at least $1-\nmo$.
\end{lemma}

\begin{proof}
	By Lemma~\ref{lemma:gssbm-concen-high-prob}, $G$ is $\C$-concentrated with probability at least $1-\nmo$. By Lemma~\ref{lemma:gssbm-concen-stable}, all graphs $G'$ (which include $G$ itself) with distance up to $\frac{c}{\epsilon}\log{n}$ from $G$ are also $\C'$-concentrated. By Lemma~\ref{lemma:gssbm-optimal}, since all graphs $G'$ are $C'$-concentrated, $SDP(G')$ always output the optimal and unique solution $\sigma^*$ with probability $1$, or $SDP(G) = SDP(G') = Y^*$ for all $G'$. It follows that $G$ is $\frac{c}{\epsilon}\log{n}$-stable under $SDP$ with probability at least $1-\nmo$.
\end{proof}

\begin{theorem}
  \label{theorem:gssbm-exact-recovery}
A graph $G$ generated by a GSSBM as in Definition~\ref{def:gssbm} 
, and $I(...) > 1$, $M_{SDP}$ with $\delta = n^{-\Omega(1)}$ exactly recovers the ground-truth community $Y^*$, i.e., $\Pr[M^{SDP}(G)\neq Y^*] = \nmo$
\end{theorem}

\begin{proof}
	Lemma~\ref{lemma:gssbm-stability-final} states the stability property of $G$ under $SDP$, i.e., $G$ is $\frac{c}{\epsilon}\log{n}$-stable under $SDP$. It also implies that $SDP(G)=Y^*$ under the effect of $\mathcal{C}$-concentration. By applying Lemma~\ref{lemma:stability-mechanism}, $\Pr[M^{SDP}(G) \neq SDP(G)] = \nmo$ or $\Pr[M^{SDP}(G) \neq Y^*] = \nmo$.

	The condition for the Theorem is that $\C'$ is a valid constant combination, or $c/\epsilon$  satisfies the condition of $c_2, c_4, c_5$. It means:

    \begin{align}
      I(a, b+\frac{2c}{\epsilon\rho_{min}}) &> 1/\rho_{min}\\
      I(b, b+ \frac{c}{\epsilon}(\frac{1}{\rho_{min}} - 1)) &> 1/\rho_{min}\\
      I(b, b+ \frac{c}{\epsilon}(\frac{2}{\rho_{min}} - 1)) &> 1/\rho_{min}
    \end{align}
\end{proof}


\end{document}